\documentclass[
 reprint,
aps,
pra,
]{revtex4-2}

\usepackage{graphics}% standard graphics specifications
\usepackage{graphicx}% Include figure files
\usepackage{longtable}% helps with long table options
\usepackage{dcolumn}% Align table columns on decimal point
\usepackage{url}% for on-line citations
\usepackage{bm}% bold math
\usepackage{hyperref}% add hypertext capabilities
\usepackage{amsmath, amssymb}
\usepackage[super]{nth}% for 1st , 2nd and 3rd numbering
\usepackage{multirow}
\usepackage{array}
\usepackage{float}
\usepackage{comment}
\usepackage[nodisplayskipstretch]{setspace} %vertical spacing setting
\usepackage{color}
\usepackage[normalem]{ulem}

\begin{document}

\preprint{APS/123-QED}
%%%%%%%%%%%%%%%%%%%%%%%%%%%%%%%%% Title %%%%%%%%%%%%%%%%%%%%%%%%%%%%%%%%%%%%%%%%%%%%
\title{Metagratings for Perfect Mode Conversion in Rectangular Waveguides:\\Theory and Experiment}% Force line breaks with \\

%%%%%%%%%%%%%%%%%%%%%%%%%%%%%%%%% Authors %%%%%%%%%%%%%%%%%%%%%%%%%%%%%%%%%%%%%%%%%%%%
\author{Vinay Kumar Killamsetty}
\email{vinay.killamsetty@gmail.com}
\author{Ariel Epstein}
 \email{epsteina@ee.technion.ac.il}
\affiliation{Andrew and Erna Viterbi Faculty of Electrical Engineering, Technion-Israel Institute of Technology, Haifa 3200003, Israel
}
%%%%%%%%%%%%%%%%%%%%%%%%%%%%%%%%% Date %%%%%%%%%%%%%%%%%%%%%%%%%%%%%%%%%%%%%%%%%%%%
\date{\today}% It is always \today, today,

%%%%%%%%%%%%%%%%%%%%%%%%%%%%%%%%% Abstract %%%%%%%%%%%%%%%%%%%%%%%%%%%%%%%%%%%%%%%%%%%%
\begin{abstract}
We present a complete design scheme, from theoretical formulation to experimental validation, exploiting the versatility of metagratings (MGs) for designing a rectangular waveguide (RWG) $\mbox{TE}_{10}$ - $\mbox{TE}_{20}$ mode converter (MC).
MG devices, formed by sparse periodically positioned polarizable particles (meta-atoms), were mostly used to date for beam manipulation applications.
In this paper, we show that the appealing diffraction engineering features of the MGs in such typical free-space periodic  scenarios can be utilized to efficiently mould fields inside waveguides (WGs).
In particular, we derive an analytical model allowing harnessing of the MG concept for realization of perfect mode conversion in RWGs.
Conveniently, the formalism considers a printed-circuit-board (PCB) MG terminating the RWG, operating as a reflect-mode MC. Following the typical MG synthesis approach, the model directly ties the meta-atom position and geometry with the modal reflection coefficients, enabling resolution of the detailed fabrication-ready design by enforcement of
the functionality constraints: elimination of the fundamental $\mbox{TE}_{10}$ reflection and power conservation (passive lossless MG). This reliable semianaltyical scheme, verified via full-wave simulations and laboratory measurements, establishes a simple and efficient alternative to common RWG MCs, typically requiring challenging deformation of the WG designed through time-consuming full-wave optimization. In addition, it highlights the immense potential MGs encompass for a wide
variety of applications beyond beam manipulation.

%\begin{description}
%\end{description}
\end{abstract}

%\keywords{Suggested keywords}% Use showkeys class option if keyword
                              % display desired
\maketitle

%\tableofcontents

%%%%%%%%%%%%%%%%%%%%%%%%%%%%%%%%% Introduction %%%%%%%%%%%%%%%%%%%%%%%%%%%%%%%%%%%%%%%%%%%%
\section{\label{sec:level1}Introduction\protect}
Metasurfaces (MSs) are ultrathin configurations of closely-spaced subwavelength scatterers (meta-atoms) with geometries exhibiting spatially-varying properties,
engineered based on generalized sheet transition conditions (GSTCs) \cite{kuester2003averaged,glybovski2016metasurfaces}.
In recent years, these structures were used to implement highly-efficient compact field-manipulating devices, such as anomalous reflectors \cite{asadchy2015functional,asadchy2016perfect,estakhri2016wave}, beam splitters \cite{epstein2016synthesis}, anomalous refraction surfaces \cite{yu2011light, pfeiffer2013metamaterial,monticone2013full,selvanayagam2013discontinuous,epstein2016arbitrary}, polarization transformers \cite{pfeiffer2016polarization}, specialized antennas \cite{epstein2016cavity, raeker2016arbitrary, minatti2016synthesis}, advanced absorbers \cite{wakatsuchi2013waveform,ra2015thin,asadchy2015broadband}, and many more, in the microwave to optics regime.
Despite their demonstrated abilities and diverse functionalities, developing physical prototypes from the abstract surface constituents obtained following the GSTC-oriented designs \cite{tretyakov2003analytical,kuester2003averaged} with high performance figures remains quite a challenging task.
The dependence on time-consuming full-wave optimization for the meta-atom design \cite{epstein2016huygens}, the requirement for sophisticated meta-atom configurations owing to the need for bianisotropic and highly nonlocal MSs for certain
functionalities \cite{estakhri2016wave,asadchy2016perfect,epstein2016arbitrary,epstein2016synthesis},
and the demand for high-end fabrication techniques to realize multi-layered closely-packed deeply subwavelength elements led to the search for new avenues for achieving unconventional beam transformations.

In the pursuit of an alternate solution, the theory of metagratings (MGs) has emerged \cite{sell2017large,khaidarov2017asymmetric,yang2018freeform}, often relying on a comprehensive analytical scheme to perform high-efficiency anomalous reflection and refraction \cite{ra2017metagratings,memarian2017wide,chalabi2017efficient,epstein2017unveiling, wong2018perfect}.
These periodic structures are also composed of meta-atoms, but, in contrast to MSs, the polarizable particles may be arranged in a sparse (not necessarily regular) manner in space.
Being periodic, when the MG is excited by an incident plane wave, scattering to a discrete set of diffraction modes (both propagating and evanescent) takes place, consistent with Floquet-Bloch (FB) theory \cite{collin1990field}.
The desired functionality is obtained by meticulous tuning of the meta-atom distribution and detailed geometry such that the induced current on the individual elements produces an interference pattern that yields the user-prescribed coupling to the various scattered FB modes.
Since their design is not based on the homogenization approximation used for MS synthesis, MGs are not limited to subwavelength interelement distances, which gives them an added advantage of relaxed fabrication requirements without the need for densely populated designs.
Importantly, the models used for MG synthesis avoid the intermediate GSTC-based abstract design step, thus leading directly to a realistic physical configuration, readily transferred to manufacturing without excessive full-wave optimizations.

\begin{figure*}[htb]
\includegraphics[scale=.4]{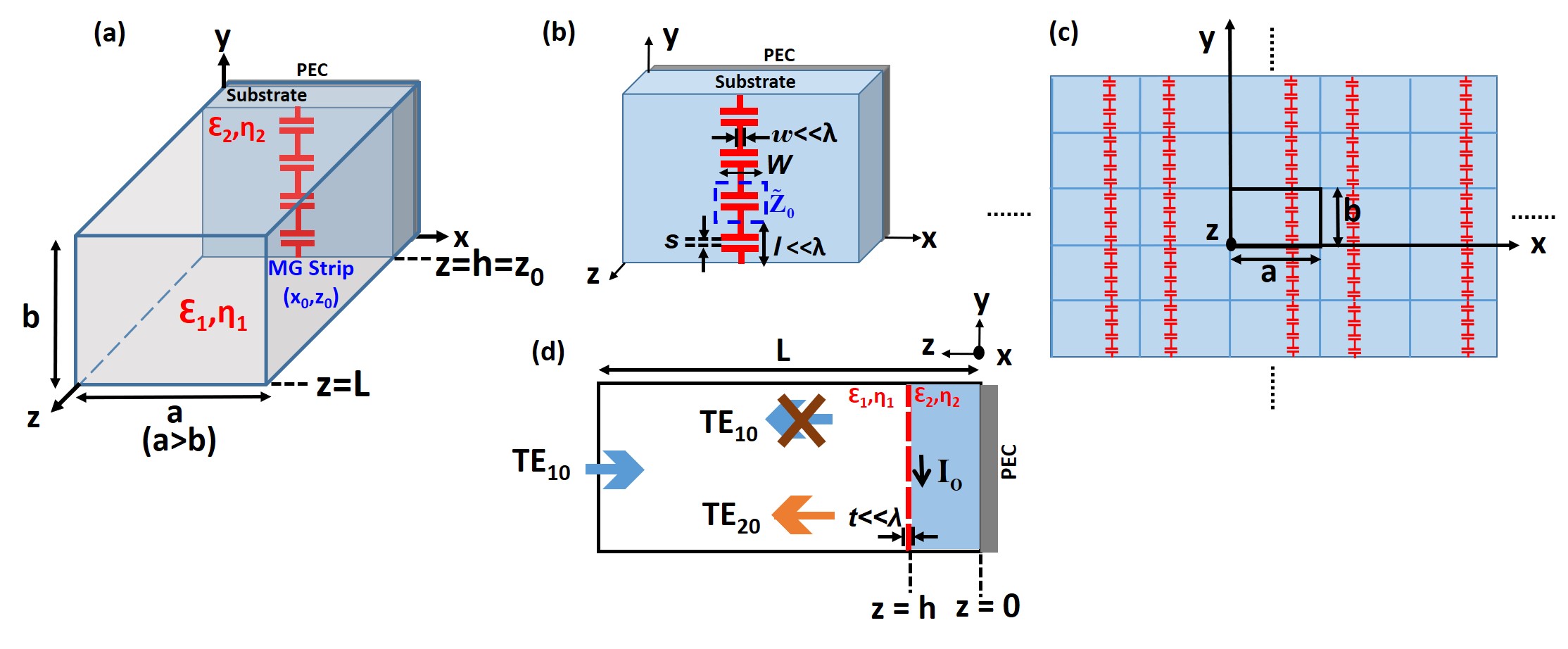}% Here is how to import EPS art
\caption{\label{fig:wide1} Schematic of the proposed RWG $\mbox{TE}_{10}$ - $\mbox{TE}_{20}$ MC configuration.
(a) Illustration of the MC, comprised of a dual-mode metallic RWG terminated with a PEC-backed dielectric substrate.
(b) Perspective view of the MG itself, featuring a capacitively-loaded conducting MG strip of thickness $t\ll \lambda$.
(c) Image theory point of view: the MC configuration forms an infinite array of images, due to the presence of (the RWG) metallic boundaries along the $\widehat{yz}$ and $\widehat{xz}$ planes at $x=0,a$ and $y=0,b$.
(d) Side view of the MC engineered to couple the excited $\mbox{TE}_{10}$ mode to $\mbox{TE}_{20}$ mode by eliminating the $\mbox{TE}_{10}$ back-reflections. }
\end{figure*}

Indeed, the burgeoning concept of MGs has found its significance in various applications including anomalous reflection \cite{chalabi2017efficient,Rabinovich2018AnalyticalReflection,Rabinovich2018AnalyticalReflection,wong2018perfect,wong2018binary,popov2018controlling,neder2019combined,popov2019constructing,popov2019designing},
anomalous refraction \cite{khaidarov2017asymmetric, yang2018freeform, sell2018ultra, fan2018perfect, casolaro2019dynamic, RabinovichArbitrary2020, dong2020efficient},
beam splitting \cite{epstein2017unveiling},
and focusing \cite{paniagua2018metalens,kang2020efficient, RabinovichArbitrary2020}, to name a few.
Nonetheless, their exceptional field-moulding capabilities were primarily utilized for beam manipulation applications in free space.
It would be beneficial, though, to harness the appealing semianalytical synthesis methodology and the highly-efficient power coupling features to solve problems in other electromagnetic configurations and scenarios.

Waveguides (WGs) play an important role in microwave communication systems, such as antennas \cite{elsherbini2012compact}, antenna feeds \cite{gonzalez2018additive,campo2020wideband,alonso2020wideband}, filters \cite{boria2007waveguide}, for their ability to transfer power and information with low loss and high fidelity.
In the case of metallic rectangular waveguides (RWGs), $\mbox{TE}_{10}$ - $\mbox{TE}_{20}$  mode converters (MCs)
\footnote{$\mbox{TE}_{mn}$ corresponds to RWG transverse electric mode of order $m$ and $n$ along the long and short dimensions of the cross section, respectively\cite{pozar2011microwave}.}
are very important components for applications such as power splitters \cite{zhao2018novel-IEEE}, spatial power combiners \cite{belaid2004mode}, circular WG mode launchers \cite{saad1977analysis,yeddulla2009analytical,liu2016design,wang2016wideband}, and channel capacity enhancers \cite{ohana2016dielectric,li2017controlling,wang2019compact}.

The usual ways of achieving $\mbox{TE}_{10}$ - $\mbox{TE}_{20}$ mode conversion in RWGs is by using junctions, e.g., orthogonal tapered RWGs \cite{saad1977analysis}, H-bends \cite{kirilenko2006nonsymmetrical,zhang2012theoretical}, E-bends \cite{zhao2018novel-AIP,zhao2018novel-IEEE}, fin-lines \cite{belaid2004mode}, and aperture coupled orthogonal RWG sections \cite{wu2013te01,liu2016design,wang2016wideband,shu2020wideband}.
In all of these reports, the discontinuity formed by the introduced junction causes excitation of all WG modes, and the optimized junction structure performs filtering and funnelling of the desired high-order mode to the output port.
Alternatively, a combination of power splitters and combiners can be used, spatially coupling an input $\mbox{TE}_{10}$ to $\mbox{TE}_{20}$ mode at the output \cite{belaid2004mode,xu2019design,shu2020design}, again involving structural WG modifications.

Indeed, most of the reported techniques for designing $\mbox{TE}_{10}$ - $\mbox{TE}_{20}$ MCs are not based on a clear analytical methodology, but rather rely on full-wave optimization of the junction to meet the design goals. Besides the potentially demanding time and memory resources required by such optimizations, the absence of an underlying analytical formulation may pose adaptation difficulties when applied to different WG specifications or operation conditions.
%Correspondingly, they typically rely on time-consuming full-wave optimization.
Furthermore, many of the proposed solutions require challenging geometries, which could lead to fabrication inaccuracies and increased realization costs.

In this paper, we address these deficiencies by presenting a complete synthesis procedure for MG-based $\mbox{TE}_{10}$ - $\mbox{TE}_{20}$ MCs, from semianalytical theoretical formulation to experimental validation with a working prototype.
In particular, we adapt the analytical model previously developed for MGs implementing engineered diffraction in free space \cite{Rabinovich2018AnalyticalReflection} to address scattering in an over-moded RWG terminated by a MG, featuring a capacitively-loaded wire printed on a metal-backed dielectric substrate [Fig.~\ref{fig:wide1}(a)].
This RWG configuration, by virtue of its metallic boundaries along the $x$ and $y$ axes, sets up an infinite loaded-wire array scenario consequent to image theory \cite{harrington2001time}, as depicted in Fig.~\ref{fig:wide1}(c).
Correspondingly, from an electromagnetic perspective, it closely resembles the free-space anomalous reflection MG configurations considered in \cite{epstein2017unveiling,Rabinovich2018AnalyticalReflection}, coupling energy of an incoming incident wave to a desired reflected FB mode.

Based on this analogy, we extend the model presented in \cite{Rabinovich2018AnalyticalReflection} to address scattering off this MG-loaded RWG. Following the typical MG synthesis scheme, we formulate analytical constraints on the resulting scattering coefficients, guaranteeing that the synthesized MG would fully couple the incoming power of an incident $\mbox{TE}_{10}$ mode to a reflected $\mbox{TE}_{20}$ mode [Fig.~\ref{fig:wide1}(d)].
Resolving these constraints in the frame of the detailed model directly yields the complete MG configuration, including the wire position, substrate thickness, and capacitor width, required to achieve the desired functionality.
These specifications are used to fabricate a prototype MG MC using standard printed-circuit-board (PCB) manufacturing techniques, verifying the fidelity of the analytical model and the efficacy of the resulting device.

As shall be demonstrated, the presented formalism can readily accommodate a wide range of WG specifications, avoids excessive full-wave optimization, and yields a fabrication-ready layout of a compact and simple PCB MC. In addition to introducing an appealing and efficient alternative to previously proposed MCs, this work emphasizes the extensive versatility of the MG concept and design approach, expected to facilitate new solutions to challenging problems in diverse electromagnetic applications and configurations, far beyond beam manipulation.

%%%%%%%%%%%%%%%%%%%%%%%%%%%%%%%%% Figures %%%%%%%%%%%%%%%%%%%%%%%%%%%%%%%%%%%%%%%%%%%%

%%%%%%%%%%%%%%%%%%%%%%%%%%%%%%%%% Theory %%%%%%%%%%%%%%%%%%%%%%%%%%%%%%%%%%%%%%%%%%%%
\section{Theory\protect}\label{sec:Theory}

%%%%%%%%%%%%%%%%%%%%%%%% Formulation (Theory subsection) %%%%%%%%%%%%%%%%%%%%%%%%%%%%
\subsection{\label{subsec:Formulation}Formulation}
We consider a dual-mode hollow metallic RWG of width $a$ ($\lambda_{0} <a<\frac{3\lambda_{0}}{2}$), height $b$ ($a>b$) and length $L$  [Fig.~\ref{fig:wide1}(a)] terminated by a PCB MG slab, where $\lambda_{0}$ is the wavelength, defined by $\lambda_{0}=c/f$, $c$ and $f$ being the speed of light in vacuum and operating frequency, respectively. The MG is composed of a perfect-electric-conductor (PEC) backed dielectric substrate [Fig.~\ref{fig:wide1}(b)] of thickness $h$ and a printed capacitively-loaded conducting strip of width $w$ and thickness $t$ [Fig.~\ref{fig:wide1}(d)] ($w,t\ll \lambda_{0}$) centered at ($x_0$, $z_0=h$) [Fig.~\ref{fig:wide1}(a)].
Time-harmonic dependency $e^{j\omega t}$ is assumed and suppressed, with $f=\omega/\left(2\pi\right)$.
The medium inside the RWG is referred to as medium 1 (free space by default) having wavenumber $k_1 = \omega\sqrt {\varepsilon_1\mu_1}$
and wave impedance $\eta_1 = \sqrt {\mu_1/\varepsilon_1}$, while the dielectric substrate of the PCB MG (medium 2) is characterized by $k_2 = \omega\sqrt {\varepsilon_2\mu_2}$ and $\eta_2 = \sqrt {\mu_2/\varepsilon_2}$; $\varepsilon_m$ and $\mu_m$ are, respectively, the permittivity and permeability of the $m$th medium.
The MG strip is periodically loaded with printed capacitors of width $W$ having a gap $s$ between the stubs.
These loads repeat along the $y$ axis with a deep sub-wavelength periodicity $l \ll \lambda_{0}$, allowing us to treat the strips as being uniformly loaded with an impedance per unit\ length $\widetilde Z_0$ \cite{tretyakov2003analytical, ikonen2007modeling}.
We excite the structure with $\mbox{TE}$ waves travelling in the positive $z$-direction $(E_{z} =0)$.
As the incident fields do not vary in the $y$-direction and the properties of the MG strip are homogenized in the $y$-direction, the total field in the RWG can be assumed uniform along $y$ $(\frac{\partial}{\partial y} = 0)$; thus, the problem effectively becomes a 2-D one \cite{epstein2017unveiling, Rabinovich2018AnalyticalReflection}.

Our goal is to utilize this structure to fully convert an incoming $\mbox{TE}_{10}$ (fundamental) mode into a $\mbox{TE}_{20}$ reflected mode.
To this end, the available degrees of freedom (DOFs) of the device, namely, the position of the MG strip ($x_0$, $h$) and the load impedance $\widetilde Z_0$, have to be set such that the coupling to reflected $\mbox{TE}_{10}$ mode is suppressed, while the incident power is funnelled to the reflected $\mbox{TE}_{20}$ mode in its entirety [Fig.~\ref{fig:wide1}(d)].
Correspondingly, following the typical MG synthesis approach \cite{ra2017metagratings,epstein2017unveiling,Rabinovich2018AnalyticalReflection,RabinovichArbitrary2020}, we first address the scattering problem via modal analysis, deriving analytical expressions for the coupling coefficients to the various WG modes. This is done by evaluating separately the contribution associated with the external field, calculated in the absence of the capacitively-loaded strip (Section \ref{subsubsec:ExternalFields}), and the fields generated by the (yet to be evaluated) current $I_0$ induced on the MG conductors due to the incident fields (Section \ref{subsubsec:InducedCurrent}). Once we establish these expressions, constraints can be formulated to guarantee the desired functionality (Sections \ref{subsec:Elimination of fundamental mode} and \ref{subsec:Passive and Lossless Condition}), the resolution of which would yield the strip position and load impedance required for implementing the MC (Section \ref{subsec:Distributed Load Impedance}).

%%%%%%%%%%%%%%%%%%%%%%%%%%%%%%% External Fields contribution %%%%%%%%%%%%%%%%%%%%%%%%%%%%%%%%
\subsubsection{External field contribution}\label{subsubsec:ExternalFields}
As mentioned above, we begin our analysis by evaluating the external field contribution, comprised of the fields reflected from the PEC-backed dielectric substrate in the absence of the loaded MG strip.
We excite the device with a fundamental RWG mode ($\mbox{TE}_{10}$) whose electric field takes the form
\begin{equation}\label{eq:incidentfields}
    E_y^{\mathrm{in}}(x,z)
    =E_{0}\sin(\frac{\pi}{a}x)e^{j\beta_{1,1}z}
\end{equation}
where $\mbox{E}_{0}$ is the amplitude of the incident $\mbox{TE}_{10}$ mode and $\beta_{n,m} = \sqrt{(k_m)^2-(\frac{n\pi}{a})^2}$  represents the longitudinal wavenumber of the $n$th-order guided mode in the $m$th medium.

Due to the interaction with the shorted dielectric slab, the incident field will undergo multiple reflections. Correspondingly, upon imposing the boundary conditions (tangential field continuity at $x=h$ and vanishing tangential electric field at $z=0$), the total external field contribution everywhere in $m$th medium $E^{\mathrm{ext}}_{{n,m}}(x,z)$ can be formulated as
\begin{equation}\label{eq:externalfields}
    \begin{array}{ll}
    E^{\mathrm{ext}}_{{1,1}}(x,z) = E_0\sin(\frac{\pi}{a}x)
    (e^{j\beta_{1,1}z}
    +R_1e^{-j\beta_{1,1}(z-2h)}) & |z|>h \\\\
    E^{\mathrm{ext}}_{{1,2}}(x,z) =E_0\sin(\frac{\pi}{a}x)
    (1+R_1)\frac{\sin(\beta_{1,2}z)}{\sin(\beta_{1,2}h)}
    e^{j\beta_{1,1}h} & |z|<h
    \end{array}
\end{equation}
where $R_n$ is the effective reflection coefficient of the $n$th guided mode from the boundary at $z=h$, given by
\begin{equation}\label{reflection and transmission}
\begin{split}
&R_n = \frac{j\gamma_n\tan(\beta_{n,2}h)-1}{j\gamma_n\tan(\beta_{n,2}h)+1} \\
\end{split}
\end{equation}
with $\gamma_n$ being the wave impedance ratio for the $n$th RWG mode, defined as $\gamma_n = {Z_{n,2}}/{Z_{n,1}}$. Here $Z_{n,m}$ is the TE wave impedance of the $n$th mode in the $m$th medium\textcolor{red}{,} given as $Z_{n,m}=k_m\eta_m/\beta_{n,m}$.

%%%%%%%%%%%%%%%%%%%%%%%%%% Induced currents contribution %%%%%%%%%%%%%%%%%%%%%%%%%%%%%%%%%%%%
\subsubsection{Induced currents contribution}\label{subsubsec:InducedCurrent}
After computing the contribution from the external fields following Eqs. (\ref{eq:externalfields}) and (\ref{reflection and transmission}), we proceed to assess the fields generated by the secondary source, namely, the current $I_0$ induced on the capacitively-loaded MG strip.
Since the dimensions of the MG strip are considered to be very small compared to the wavelength ($t,w \ll \lambda$), it can be modelled well by an infinitesimal line source at $(x_0,h)$.
This substitution allows us to invoke the WG Green's function to evaluate the associated secondary fields
\cite{lewin1975theory,leviatan1983single}, taking into consideration the boundary conditions discussed in Section \ref{subsubsec:ExternalFields}, which apply here as well.

Subsequently, for a current distribution  $\vec{J}\left(x,z\right)=I_0\delta\left(x-x_0\right)\delta\left(z-h\right)\hat{y}$, the fields in the $m$th medium can be expressed as a modal sum $E_{{m}}^{\mathrm{strip}}\left(x,z\right)=\sum_{n=1}^\infty E_{{n,m}}^{\mathrm{strip}}\left(x,z\right)$ where the fields associated with the $n$th WG mode can be generally written as \cite{lewin1975theory,leviatan1983single}
\begin{equation}\label{eq:general field eq for stacked substrate}
\begin{split}
    E_{{n,m}}^{\mathrm{strip}} =
   &\left(-\frac{2I_0}{a}
    Z_{n,m}\right)
    \sin(\frac{n\pi}{a}x_0)\sin(\frac{n\pi}{a}x)\\
    &\cdot(A_{n,m}e^{j\beta_{n,m}z}+B_{n,m}e^{-j\beta_{n,m}z})
\end{split}
\end{equation}
with $A_{n,m}$ and $B_{n,m}$, respectively, standing for the amplitudes of the forward and backward propagating waves of the $n$th RWG mode in the $m$th medium.
These amplitudes are determined by applying the PEC boundary conditions at $z=0$ ($A_{n,2}=-B_{n,2}$), the radiation condition at $z\rightarrow\infty$ ($B_{n,1}=0$), and the source condition at $z=h$, formulated using impulse balance techniques as \cite{RabinovichArbitrary2020}
\begin{equation}\label{eq:BoundaryConditions-1}
\begin{split}
%&E_{y_{2,n}}^{strip}\Big|_{z\rightarrow 0^{+}} =0 \\
&E_{{n,2}}^{\mathrm{strip}}\Big|_{z\rightarrow h^{-}} = E_{{n,1}}^{\mathrm{strip}}\Big|_{z\rightarrow h^{+}} \\
&H_{{n,2}}^{\mathrm{strip}}\Big|_{z\rightarrow h^{-}} - H_{{n,1}}^{\mathrm{strip}}\Big|_{z\rightarrow h^{+}}= \frac{I_0}{a}\sin(\frac{n\pi}{a}x)\sin(\frac{n\pi}{a}x_0)
\end{split}
\end{equation}
Upon imposing these conditions on the modal expansion, we arrive at the following expressions for the fields associated with the current-carrying strips when the observation point is in medium $1$ ($z \geq h$)
\begin{equation}\label{eq:secondarySource in stacked freespace}
\begin{split}
   & E_{{1}}^{\mathrm{strip}}(x,z) = \sum_{n=-\infty}^\infty E_{{n,1}}^{\mathrm{strip}}(x,z)\\
   &\!=\!-\frac{2I_0}{a}\!\!
    \sum_{n=1}^\infty
    \!Z_{n,1}(1+R_{n})
    \sin(\frac{n\pi}{a}x_0)\sin(\frac{n\pi}{a}x)e{^{-j\beta_{n,1}(z-h)}}
\end{split}
\end{equation}
and medium $2$
\begin{equation}\label{eq:secondarySource in stacked susbtrate}
\begin{split}
   & E_{{2}}^{\mathrm{strip}}(x,z) = \sum_{n=-\infty}^\infty E_{{n,2}}^{\mathrm{strip}}(x,z)\\
   & \!=\!-\frac{2I_0}{a}
    \sum_{n=1}^\infty \left\lbrace
    \begin{array}{l}
        Z_{n,2}(1+R_{n})
    \\
    \sin(\dfrac{n\pi}{a}x_0)\sin(\dfrac{n\pi}{a}x)\dfrac{1}{\gamma_n}\dfrac{\sin(\beta_{n,2}z)}{\sin(\beta_{n,2}h)}
    \end{array}
    \right\rbrace
\end{split}
\end{equation}
where $R_n$ is the reflection coefficient defined in Eq. \eqref{reflection and transmission}.

Finally, the entire fields inside the WG can be evaluated by adding the
external fields of Eq. (\ref{eq:externalfields}) to the modal contributions resulting from the current induced on
the MG strip [Eq\textcolor{red}{s}. (\ref{eq:secondarySource in stacked freespace}) and (\ref{eq:secondarySource in stacked susbtrate})], yielding
\begin{equation}\label{eq:Total fields}
E_{{m}}^{\mathrm{tot}}(x,z)=E^{\mathrm{ext}}_{{1,m}}(x,z)+
\sum_{n=1}^{\infty} E_{{n,m}}^{\mathrm{strip}}
\end{equation}
%%
%%%%%%%%%%%%%%%%%%%%%%%%%% Elimination of TE10 mode %%%%%%%%%%%%%%%%%%%%%%%%%%%%%%%%%%%%
\subsection{Suppressing $\mbox{TE}_{10}$ Reflection}\label{subsec:Elimination of fundamental mode}
As denoted in Section \ref{subsec:Formulation}, we operate the RWG in the dual-mode regime, allowing only two propagating modes inside the RWG: $\mbox{TE}_{10}$ and $\mbox{TE}_{20}$ (the rest of the modes will be evanescent). These operating conditions are expected to enable realization of perfect mode conversion with a single meta-atom, similar to perfect anomalous reflection in free-space MG configurations \cite{ra2017metagratings,epstein2017unveiling,Rabinovich2018AnalyticalReflection}.
Analogously, to fully convert the incident $\mbox{TE}_{10}$ mode [Eq. (\ref{eq:incidentfields})] into an outgoing $\mbox{TE}_{20}$ mode, we should first demand that the reflected external $\mbox{TE}_{10}$ mode fields [$R_1$ term of Eq. (\ref{eq:externalfields})] should destructively interfere with the fields scattered into $\mbox{TE}_{10}$ mode due to the currents induced on the MG [$n = 1$ term in Eq. (\ref{eq:secondarySource in stacked freespace})], such that no power is coupled to the spurious $\mbox{TE}_{10}$ reflection [Fig. \ref{fig:wide1} (d)].
Formally, the elimination of coupling to the $\mbox{TE}_{10}$ mode is obtained by requiring that the corresponding term in Eq. (\ref{eq:Total fields}) would vanish, which implies that the geometry of the meta-atom at $(x_0,h)$ should be tuned such that the induced currents on it would satisfy %\textcolor{red}{(compare with \cite{Rabinovich2018AnalyticalReflection})} \textcolor{blue}{\textbf{[There is a correspondence, right?]}}
\begin{equation}\label{eq:Current}
I_0 = E_0\frac{e^{j\beta_{1,1}h}}
{2\left(\frac{Z_{1,1}}{a}\right)\sin(\frac{\pi}{a}x_0)}
\left(\frac{R_1}{1+R_1}\right)
\end{equation}

%%
 %%%%%%%%%%%%%%%%%%%%%%%%Passive and Lossless Condition%%%%%%%%%%%%%%%%%%%%%%%%%%%%
\subsection{Perfect Mode Conversion}\label{subsec:Passive and Lossless Condition}
While the condition of Eq. \eqref{eq:Current} ensures elimination of spurious reflection into the fundamental $\mbox{TE}_{10}$ mode, it does not guarantee that it is possible to excite the required current $I_0$ derived therein without incurring loss or introducing gain into the system.
Hence, in order to allow implementation of the mode conversion functionality using a passive and lossless configuration, thus avoiding undesired absorption or complicated active elements, an additional constraint should be enforced, requiring total power conservation \cite{ra2017metagratings,epstein2017unveiling,Rabinovich2018AnalyticalReflection}.
Specifically, we require that the total net real power crossing a given cross-section of the waveguide at $z>h$ must vanish, namely,
\begin{equation}\label{eq:Power Conservation definition}
{P_z(z)} = \frac{1}{2}
\int_{0}^{a}\int_{0}^{b}
\Re{\{E_{{1}}^{\mathrm{tot}} \times H_{{1}}^{\mathrm{tot}^*}\}}dxdy = 0
\end{equation}
Substituting the electric fields of Eq. (\ref{eq:Total fields}) and the associated magnetic fields, evaluated as $H_{{1}}^{\mathrm{tot}} =
 \frac{1}{jk_1\eta_1}
\frac{\partial}{\partial z}
E_{{1}}^{\mathrm{tot}}
$, into Eq. (\ref{eq:Power Conservation definition}) while taking Eq. (\ref{eq:Current}) into account yields the condition for $\mbox{TE}_{10}$-$\mbox{TE}_{20}$ perfect mode conversion, reading
%
%% Equation
\begin{equation}\label{eq:condition for perfect cconversion}
\rho\triangleq
\frac{\left|1+R_1\right|^2}{\left|1+R_2\right|^2} -
\left[
\frac{\beta_{1,1}\sin^2\left( \frac{2\pi}{a}x_0 \right)}
{\beta_{2,1}\sin^2\left( \frac{\pi}{a}x_0 \right)}
\right]
 = 0
\end{equation}
where we defined the deviation from the perfect mode conversion condition as $\rho$. This result forms a nonlinear equation for the MG strip coordinates ($x_0$, $h$), the solutions of which correspond to positions where suitable passive and lossless loaded wires can be placed to achieve the desired functionality.

%%%%%%%%%%%%%%%%%%%%%%%%%Evaluating Required Distributed Impedance%%%%%%%%%%%%%%%%%%%
\subsection{Distributed Load Impedance}\label{subsec:Distributed Load Impedance}
Choosing one of the valid MG strip positions ($x_0$, $h$) solving Eq. \eqref{eq:condition for perfect cconversion} and substituting it into Eq. \eqref{eq:Current}, yields the induced current $I_0$ required to realize perfect mode conversion without disturbing the power balance in the system.
However, we recall that our goal is to engineer the passive MG load geometry such that this current will be developed self-consistently on the MG as a response to the incident field, without using any additional external sources.
The first step in this direction would be to assess the (per-unit-length) load impedance $\widetilde{Z}_0$ [Fig.\ref{fig:wide1}(b)] that the printed capacitor should form to this end.
Following \cite{epstein2017unveiling,Rabinovich2018AnalyticalReflection,rabinovich2019experimental,RabinovichArbitrary2020}, we invoke Ohm's law \cite{tretyakov2003analytical}, relating the total fields over the MG strip [Eq. (\ref{eq:Total fields})] at ($x_0$, $h$) to the current induced over it via the distributed load impedance, leading to
\begin{equation}\label{eq:ohms law}
\begin{split}
\widetilde{Z}_0I_0
%& = E_{y_1}^{Tot}(x_0,z_0)\\ %(\Longrightarrow \textbf{Ohm's law}) \\
& =E^{\mathrm{ext}}_{{1,1}}(x_0,h)+ \underbrace{\sum_{n=1}^{\infty}E_{{n,1}}^{\mathrm{strip}}(x\rightarrow x_{0},z\rightarrow h)}_{E_{{1}}^{\mathrm{strip}}}
\end{split}
\end{equation}

We further distinguish between the secondary fields that would have been generated by the current-carrying strips in the absence of the PEC-backed substrate $E_{{1}}^{\mathrm{source}}\left(x,z\right)$ and the ones produced due to the image formed by the reflecting termination $E_{{1}}^{\mathrm{image}}\left(x,z\right)$, defining
\begin{equation}\label{eq:source and image fields}
\begin{split}
&E_{{1}}^{\mathrm{strip}}(x,z)
=E_{{1}}^{\mathrm{source}}(x,z)+E_{{1}}^{\mathrm{image}}(x,z)\\
& \!=\!
-\frac{2I_0}{a}
\sum_{n=1}^\infty
Z_{n,1}
\sin\bigg(\frac{n\pi}{a}x\bigg)\sin\bigg(\frac{n\pi}{a}x_0\bigg)e^{-j\beta_{n,1}(z-h)}\\
&\!-\!\frac{2I_0}{a}
\sum_{n=1}^\infty
Z_{n,1}
\sin\bigg(\frac{n\pi}{a}x\bigg)\sin\bigg(\frac{n\pi}{a}x_0\bigg)R_{n}e^{-j\beta_{n,1}(z-h)}\\
\end{split}
\end{equation}
This distinction is important, since directly substituting ($x_0$, $h$) for ($x,z$) in Eq. \eqref{eq:secondarySource in stacked freespace} to evaluate the fields over the MG strip would not be feasible, due to the diverging nature of the fields from a current source at the point of origin \cite{Rabinovich2018AnalyticalReflection}.
However, once we separate the terms as in Eq.  \eqref{eq:source and image fields}, this singularity manifests itself only in the source-contribution summation, while the image-related terms can be safely evaluated everywhere in space.
To resolve the singularity issues in the former, we employ the technique utilized in \cite{Rabinovich2018AnalyticalReflection,RabinovichArbitrary2020},  rewriting $E_{{n,1}}^{\mathrm{source}}\left(x,z\right)$ in the form of a Hankel function summation,  subsequently isolating the divergent term from the rest of the series, and treating it separately using the flat wire approximation.
Finally, the distributed load impedance of the MG strip required to induce the current $I_0$ given by Eq. (\ref{eq:Current}) in response to the external WG excitation, facilitating perfect power coupling from the incident $\mbox{TE}_{10}$ mode to the reflected $\mbox{TE}_{20}$ mode, is found as (see Appendix \ref{Appendix})
\begin{equation}\label{eq:source and image fields regular}
\begin{split}
&E_{{1}}^{\mathrm{strip}}(x\rightarrow x_{0},z\rightarrow h) =
\\
&- \left(\frac{k_1\eta_1}{2a}\right)
\left[
\begin{split}
& 4\sum_{n=1}^{\infty}
\left\{ \frac{\sin^2\left(\frac{n\pi}{a}x_0\right)}{\beta_{n,1}}
 - j\left(\frac{2a}{n\pi}\right)
\right\}\\
& +j\left(\frac{2a}{\pi}\log\frac{4a}{\pi w}
\right)
\end{split}
\right]\\
&-\frac{2I_0}{a}
\sum_{n=1}^\infty
Z_{n,1}R_{n}
\sin^2\bigg(\frac{n\pi}{a}x_{0}\bigg)
\end{split}
\end{equation}

Upon substituting Eqs. (\ref{eq:Current}), (\ref{eq:condition for perfect cconversion}), and \eqref{eq:source and image fields regular} into Eq. \eqref{eq:ohms law}, the load impedance per unit length required to completely couple the incident $\mbox{TE}_{10}$ into reflected $\mbox{TE}_{20}$ can be explicitly written as
\begin{equation}\label{eq:Impedance}
\begin{split}
&\widetilde{Z}_0 =
-j
\left(
4\frac{Z_{2,1}|1+R_{2}|^{2}}{a}
\right)
\\
&\cdot
\bigg[
\left(
\frac{1}{\gamma_{1}\tan{(\beta_{1,2}h)}}
+\frac{1}{\gamma_{2}\tan{(\beta_{2,2}h)}}\right)
\sin^2\Big(\frac{2\pi}{a}x_{0}\Big)
\bigg]
\\
&+ j \left(\frac{k_1\eta_1}{2a}\right)
\left[
\begin{split}
& \sum_{n=1}^{\infty}
 \left(
 \frac{j2a}{n\pi}
 \right)
 +\left(
 \frac{2a}{\pi}\log\frac{4a}{w\pi}
\right)\\
& - \sum_{n=3}^{\infty}
4\frac{(1+R_n)}{\alpha_{n,1}}
\sin^2\left(\frac{n\pi}{a}x_0
\right)
\end{split}
\right]
\end{split}
\end{equation}
where $\alpha_{n,m}=-j\beta_{n,m}$. As can be seen, the resulting load impedance is indeed purely imaginary, in consistency with the implemented power conservation requirement, leading to a passive and lossless MG MC.

Overall, for given WG configuration, substrate permittivity, and input $\mbox{TE}_{10}$ mode, the device specifications required to obtain perfect mode conversion, namely, the MG strip horizontal position $x_0$, the substrate thickness $h$, and the distributed load impedance $\widetilde{Z}_0$, can be extracted via Eqs. (\ref{eq:condition for perfect cconversion}) and (\ref{eq:Impedance}). The last step towards practical realization, replacing the lumped load impedance with a judiciously designed printed capacitor geometry, will be addressed following the methodology presented in \cite{Rabinovich2018AnalyticalReflection}, as shall be discussed and illustrated in detail in subsequent sections, finalizing the synthesis procedure.

%%%%%%%%%%%%%%%%%%%%%%%%%%% Results and Discussion %%%%%%%%%%%%%%%%%%%%%%%%%%%%%%%%%%
\section{Results and Discussion\protect}\label{sec:results and discussion}
%%%%%%%%%%%%%%%%%%%%%%%%%%% Design and Full-wave simulation %%%%%%%%%%%%%%%%%%%%%%%%%%%%%%%%%%
%
\subsection{Design and full-wave verification}\label{subsec:MG simulation}
%

%% Figure
\begin{figure*}[htb]
\includegraphics[scale=0.5]{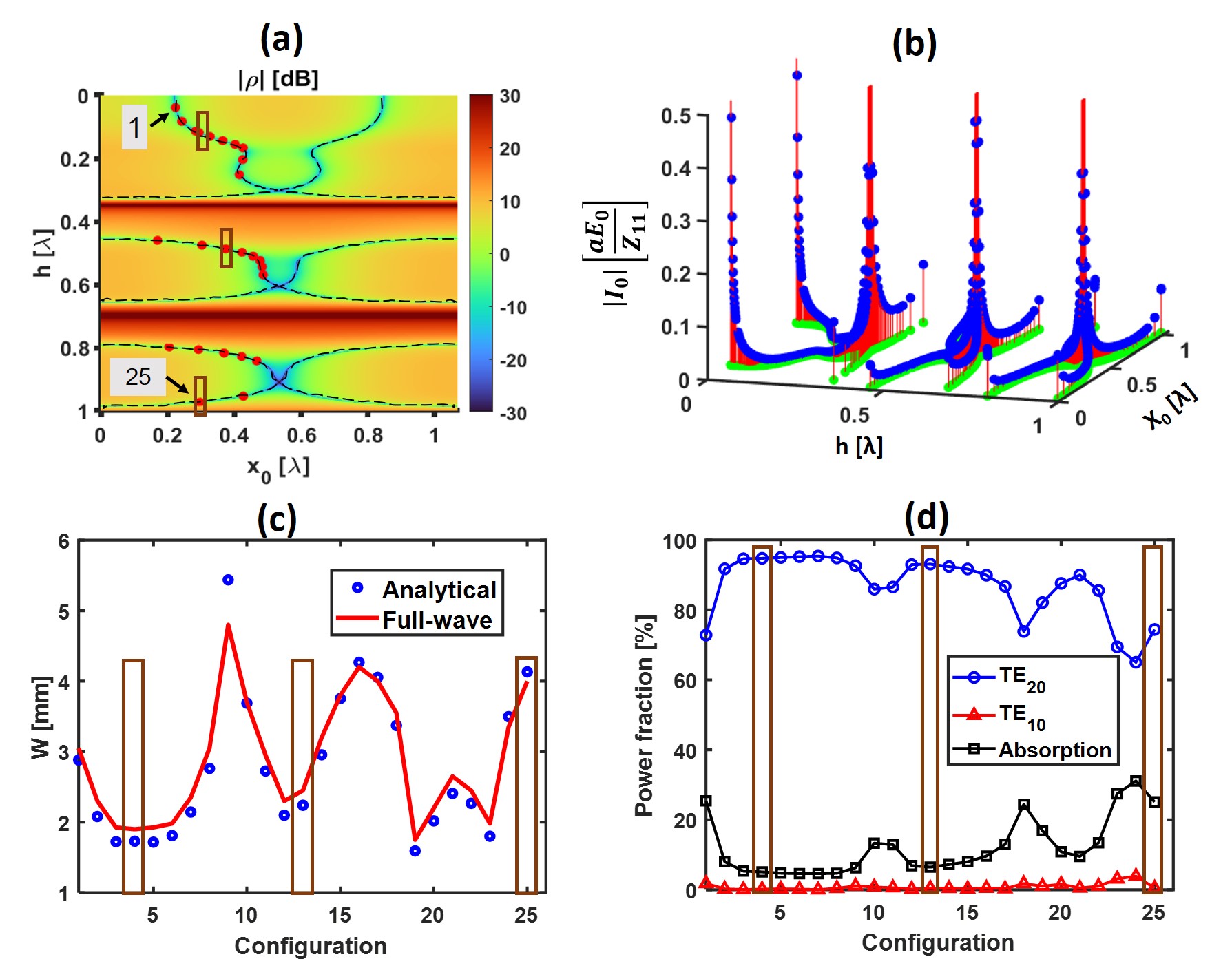}% Here is how to import EPS art
\caption{\label{fig:Widths of the capacitor}
(a) Deviation from the perfect MC condition $\left|\rho\right|$, as defined in Eq. \eqref{eq:condition for perfect cconversion}, as a function of position of the MG strip $(x_0,h)$, in dB scale. Dashed black lines denote the solution branches, indicating locations leading to valid (passive lossless) MC designs.
The 25 red points, numbered $1-25$, correspond to different possible RWG MC design configurations, subsequently characterized in (c) and (d).
(b) Induced current magnitude required to eliminate the reflected $\mbox{TE}_{10}$ mode [Eq. (\ref{eq:Current})] as a function of the strip coordinates, for valid positions along the solution branches of (a).
(c) Load capacitor widths required to realize perfect mode conversion, for each of the MC configurations marked in (a); analytically-predicted values [Eq. (\ref{eq:Width})] (blue circles) are compared to the optimal widths obtained via full-wave simulations (solid red line).
(d) Mode conversion efficiency (blue circles), spurious reflection (red triangles), and absorption (black squares), as recorded by full-wave simulations for the 25 MC configurations marked in (a) and (c). Brown rectangles denote configurations $\#4, \#13,$ and $\#25$ in (a), (c), and (d)
}
\end{figure*}
To illustrate and validate our methodology, we follow the semianalytical synthesis scheme prescribed in Section \ref{sec:Theory} to design a $\mbox{TE}_{10}$ - $\mbox{TE}_{20}$ MC operating at $f_0=14$ GHz (wavelength of $\lambda_0=21.413$ mm).
The system configuration is composed of a WR-90 hollow metallic RWG, a PEC-backed Rogers RT/Duroid 6002 dielectric substrate (permittivity $\varepsilon_2 = 2.94\varepsilon_0$ and loss tangent $\tan\delta = 0.0012)$, the top layer of which features patterned copper implementing the loaded-strip MG (see Fig. \ref{fig:wide1}).
At the operating frequency, the standard WR-90 dimensions
$a \times b$ = 22.86 $\times$ 10.16 $\text{mm}^2 \; (= 1.067\lambda_0 \times 0.474\lambda_0)$ \cite{retangularwaveguidestandards} support only two propagating modes, as required (see Section \ref{subsec:Elimination of fundamental mode}).
The MG strip is composed of copper traces (conductivity $\sigma=$ 5.8$\times10^7$ S/m) of width $w$ = 10 mil = 254 $\mu$m and thickness  $t$ = 18 $\mu$m.
The periodicity of the impedance ($\widetilde{Z}_0$) loading the MG strip along the $y$-axis is set as $l = 0.118\lambda_0 \ll \lambda_0$,  allowing homogenization along this axis (see Section \ref{subsec:Formulation}).

The first step in the design process is to identify these locations ($x_0,h$) for the MG strip that would facilitate mode conversion with unitary efficiency via a reactive (passive lossless) load, as derived in Eq. \eqref{eq:condition for perfect cconversion}.
To this end, we plot in Fig. {\ref{fig:Widths of the capacitor}}(a) the deviation from the perfect mode conversion condition $\left|\rho\right|$ for the suitable range of coordinates $x_0 \in (0,a)$,  $z_0 \in (0,\lambda]$ [Fig. \ref{fig:wide1}(a)], in dB scale.
Being a nonlinear overdetermined equation, Eq. \eqref{eq:condition for perfect cconversion} yields multiple solution branches, corresponding to minima of $\left|\rho\right|$.
These branches, marked with black dashed lines in Fig. \ref{fig:Widths of the capacitor}(a), are clearly symmetric with respect to $x_0=a/2$, manifesting the symmetry of the RWG configuration and mode conversion functionality.

In principle, all the points that lie on the denoted solution branches serve as valid coordinates for the MG strip.
However, as highlighted previously in \cite{epstein2017unveiling,Rabinovich2018AnalyticalReflection}, using some of these locations for the MG design may give rise to significant losses in practice.
As discussed in Section \ref{subsec:Elimination of fundamental mode}, to enable unitary mode conversion it is required that the reflected external $\mbox{TE}_{10}$ field would destructively interfere with the fundamental FB mode generated by the current induced on the MG strip, while radiation from the latter is exclusively coupled to the $\mbox{TE}_{20}$.
Nonetheless, some allegedly valid solutions to Eq. \eqref{eq:condition for perfect cconversion} may place the MG strip in such coordinates ($x_0,h$) where radiation capability to the $\mbox{TE}_{10}$ is highly suppressed, due to destructive interference with the images formed by the RWG metallic walls; such inability to radiate would prevent cancelling the spurious term in the external field expression [Eq. \eqref{eq:externalfields}].
Furthermore, some of these locations, e.g. close to the RWG center $x_0\rightarrow a/2$, prohibit coupling from the excited MG strip to the $\mbox{TE}_{20}$ mode due to vanishing overlap integrals; once again, placing a scatterer at these points would impede the desired mode conversion.

For such locations on the solution branches, the current $I_0$ that needs to be induced on the MG strips as per Eq. (\ref{eq:Current}) to facilitate the required functionality would be extremely large, attempting to overcome the poor coupling to either of the relevant modes.
However, physical realizations of the designed MG-based MC, which inevitably include finite conductor losses (even if small), would result in substantial power dissipation in the MG strip, which might greatly reduce the device efficiency \cite{epstein2017unveiling,Rabinovich2018AnalyticalReflection}. Plotting in Fig. \ref{fig:Widths of the capacitor}(b) these required currents as a function of $(x_0,h)$ along the solution branches found in Fig. \ref{fig:Widths of the capacitor}(a) indeed reveals a substantial increase in $I_0$ near the RWG walls or close to its center.
Hence, to retain the high performance figures predicted for the MG MC also in practical scenarios, one should locate the MG strip away from these problematic points, which are more prone to losses.

Once a suitable location for the MG strip ($x_0,h$) is found, we substitute it into Eq. (\ref{eq:Impedance}) to retrieve the (reactive) load impedance $\widetilde{Z}_0$ required to facilitate excitation of the current $I_0$ prescribed by Eq. (\ref{eq:Current}), in response to the incident $\mbox{TE}_{10}$ mode. %guaranteed by the formulation to be reactive.
For the considered scenario and functionality, with the chosen deep-subwavelength strip width $w$, this required reactance turns out to be capacitive, thus can be implemented using the printed capacitor geometry described in Fig. \ref{fig:wide1} \footnote{In case inductive loading is required, other PCB-compatible geometries, such as meander lines, may be used \cite{popov2019designing}.}.
The lumped capacitor $C_{\mathrm{load}}$ value corresponding to the evaluated impedance $\widetilde{Z}_0$ at the operating frequency is readily determined via
\begin{equation}\label{eq:Capcitor}
C_{\mathrm{load}}=
-\frac{1}{2\pi l f_0\Im({\widetilde{Z}_0})}
\end{equation}
To realize this load capacitance using printed copper traces, we utilize the method presented in \cite{epstein2017unveiling,Rabinovich2018AnalyticalReflection}, approximating the printed capacitor width $W$ required to implement a given $C_{\mathrm{load}}$ with the geometry presented in Fig. \ref{fig:wide1} following
\begin{equation}\label{eq:Width}
W\approx
\frac{2.85K_{\mathrm{corr}}C_{\mathrm{load}}}{\varepsilon_{\mathrm{eff}}}
\left[
\frac{\mathrm{mil}}{\mathrm{fF}}
\right]
\end{equation}
where $K_{\mathrm{corr}}$ is a frequency-dependent correction factor, found to be $K_{\mathrm{corr}}=2.1$ for our configuration (i.e., at 14GHz and for $w=s=10\;\mathrm{mil}$) \footnote{The correction factor $K_{\mathrm{corr}}$ is assessed with the aid of a full-wave simulation. For a chosen reference case [configuration $\#10$ of Fig. \ref{fig:Widths of the capacitor}(a) in our case, we find the optimal capacitor width in Ansys HFSS and use this value to calibrate $K_{\mathrm{corr}}$ such that Eq. \eqref{eq:Width} would yield the same result \cite{epstein2017unveiling}.}, and $\varepsilon_{\mathrm{eff}}=(\varepsilon_1+\varepsilon_2)/2$ is the effective (average) permittivity in the vicinity of the capacitor \cite{epstein2017unveiling, Rabinovich2018AnalyticalReflection}.
This step concludes our semianalytical design procedure, yielding a complete fabrication-ready layout for a PCB-based RWG MC.
%extracting all the required design parameters of our RWG $\mbox{TE}_{10}$ - $\mbox{TE}_{20}$ MC configuration with a rigorous semi-analytical methodology.

To validate the theoretical model, we follow the outlined scheme to synthesize different RWG MC configurations, subsequently defined and simulated in a commercial full-wave solver (Ansys HFSS).
Each configuration corresponds to a unique MG strip location, found by solving Eq. (\ref{eq:condition for perfect cconversion}).
In particular, for these verification purposes, we consider twenty-five such locations, marked with red dots on the solution branches highlighted in Fig. \ref{fig:Widths of the capacitor}(a) (numbered 1 to 25). %with red dots referenced from 1 to 25 are simulated in full-wave solver.
For each strip position $(x_0,h)$, the corresponding load capacitor widths $W$ (blue circles) are calculated via Eqs. (\ref{eq:Impedance})-(\ref{eq:Width}) and respectively plotted in Fig. \ref{fig:Widths of the capacitor}(c).

%% Figure
\begin{figure*}[htb]
\includegraphics[scale=.45]{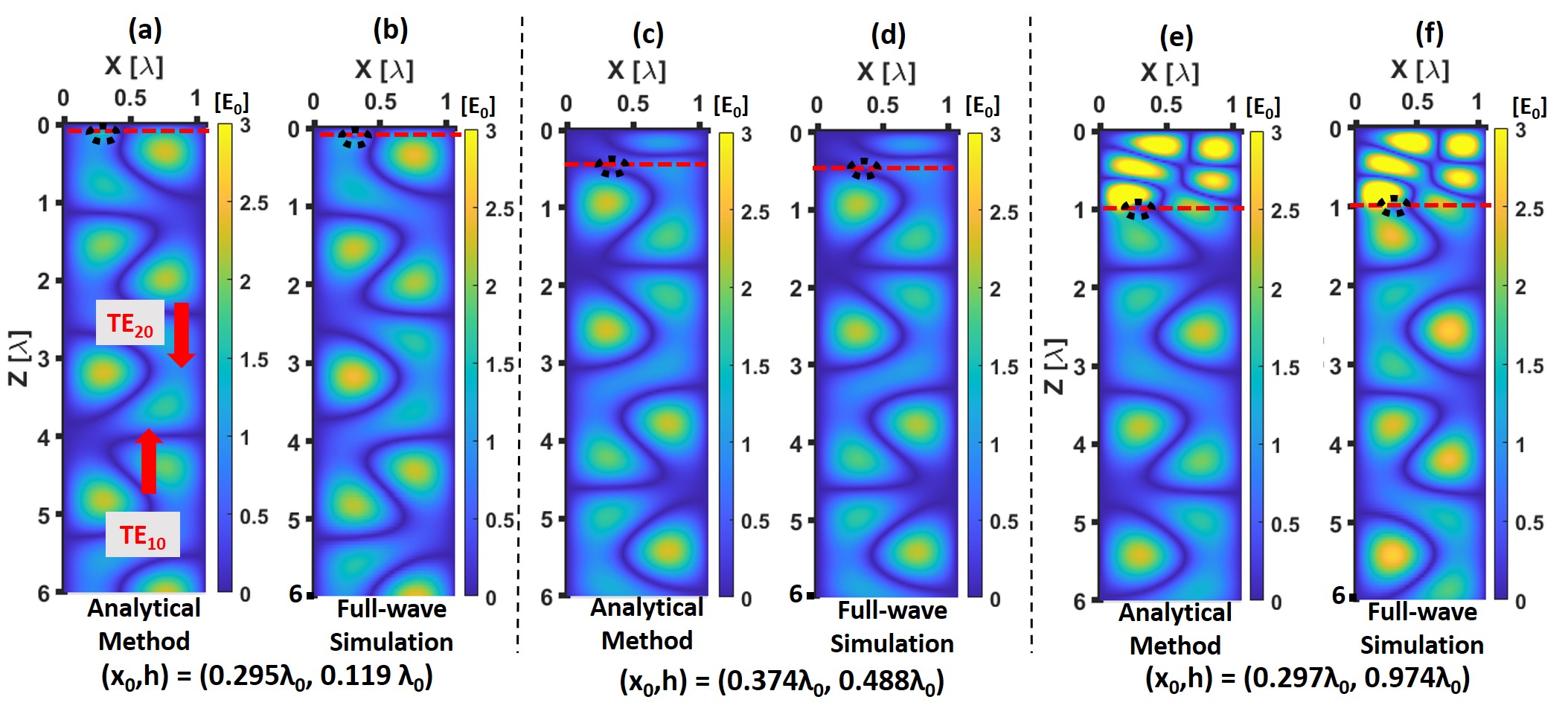}% Here is how to import EPS art
\caption{\label{fig:E-field plots}
Analytically predicted (a),(c),(e) and full-wave simulated (b),(d),(e) electric field distributions $|\Re({E_{y}^{\mathrm{tot}}(x,z))}|$ of the RWG MC configurations at the operating frequency $f=14$ GHz, corresponding to (a),(b) ($x_0,h$) =($0.295\lambda_0,0.119\lambda_0$) [configuration $\#4$ in Fig. \ref{fig:Widths of the capacitor}],
(c),(d) ($x_0,h$) =($0.374\lambda_0,0.488\lambda_0$)  [$\#13$ in Fig. \ref{fig:Widths of the capacitor}] and
(e),(f) ($x_0,h$) =($0.297\lambda_0,0.974\lambda_0$)  [$\#25$ in Fig. \ref{fig:Widths of the capacitor}].
The dashed black circle marks the position of the MG strip and the dashed red horizontal lines denote the substrate/air interface ($z=h$).
}
\end{figure*}

As a first verification step, we define the resultant PCB MG MC configurations in Ansys HFSS [Fig. \ref{fig:wide1}(a)], excite them with the fundamental $\mbox{TE}_{10}$ from the input port ($L = 6\lambda$), and conduct for each a parametric sweep to find the capacitor width value that would maximize the power coupled to the $\mbox{TE}_{20}$ mode.
The optimal widths obtained from the simulations, plotted in Fig. \ref{fig:Widths of the capacitor}(c) in solid red, indicate a very good agreement with the analytically predicted values for $W$.
This agreement highlights the fidelity of the semianalytical formalism, enabling reliable estimation of the detailed meta-atom geometry for our design without resorting to extensive full-wave optimization.

Next, we check whether these meta-atoms, when positioned at the coordinates $(x_0,h)$ prescribed by the semianalytical scheme, indeed realize the desired mode conversion functionality.
To this end, we extract from the full-wave simulations, for each of the twenty-five configurations marked in Fig. \ref{fig:Widths of the capacitor}(a), featuring the $W$ found in Fig. \ref{fig:Widths of the capacitor}(c), the relative power coupled to each of the RWG guided modes.
The simulated results are presented in Fig. \ref{fig:Widths of the capacitor}(d), showing the fraction of the incident $\mbox{TE}_{10}$ power converted to reflected $\mbox{TE}_{20}$ fields (blue circles), as required, alongside residual power remaining in the $\mbox{TE}_{10}$ mode (red triangles) or absorbed due to the finite conductivity of the printed copper traces used in simulations (black squares).
As observed, in all of the considered designs generated by our method, practically no power is coupled back to the $\mbox{TE}_{10}$ mode.
In other words, the RWG PCB MG MCs succeed in completely suppressing spurious reflections, as prescribed in Section \ref{sec:Theory}.
Due to the realistic conductor loss, at these strip locations where the induced current is predicted analytically to be higher [Fig. \ref{fig:Widths of the capacitor}(b)], one indeed notices increased absorption.
Nonetheless, even when considering these practical aspects, it is clear that Ohmic dissipation can be kept small by judicious selection of the MG working point along the solution branch, with most of the examined MG MCs exhibiting conversion efficiencies of above $90\%$.
This validates the ability of the proposed analytical scheme to yield realistic highly-efficient PCB-based MCs.

Finally, we conclude this numerical verification section by examining the scattered fields for three representative RWG MG MC configurations, encircled in Fig. \ref{fig:Widths of the capacitor}(a), (c), and (d) by brown rectangles.
For these MGs, corresponding to strip coordinates ($x_0,h$) = ($0.295\lambda_0,0.119\lambda_0$), ($0.374\lambda_0,0.488\lambda_0$), and ($0.297\lambda_0,0.974\lambda_0$) we compare in Fig. \ref{fig:E-field plots} the field snapshots as predicted by the analytical model [following Eq. \eqref{eq:Total fields} with Eq. (\ref{eq:Current})] with those recorded in HFSS, when the structure is excited by an incoming $\mbox{TE}_{10}$ wave.
These plots, showing the expected interference pattern between the incident $\mbox{TE}_{10}$ mode and the reflected $\mbox{TE}_{20}$ mode, attest once again to the close agreement between the theoretical prediction and the full-wave modelled prototype, further validating our analytical approach.
Since the synthesis method presented in Section \ref{sec:Theory}, the accuracy of which was verified via simulations in this subsection, yields detailed fabrication-ready designs, one can directly proceed to experimental validation via a suitable prototype.

\subsection{Experimental validation}\label{subsec:Experimental validation}
For experimental ratification, we choose to manufacture a prototype MG, relying on
the configuration corresponding to $(x_0,h) = (0.295\lambda_0,\;0.119\lambda_0)$ [Fig. \ref{fig:E-field plots}(a),(b); $\#4$ in Fig. \ref{fig:Widths of the capacitor}]. This is a convenient working point, since its MG strip location coincides with the thickness of a commercially available Rogers RT/duroid 6002 dielectric laminate at the operating frequency, namely, $h=2.54$ mm.
In addition, as shown in Fig. \ref{fig:Widths of the capacitor}(d), this configuration features a high conversion efficiency, with $\approx 95\%$ of the input power coupled to the higher-order $\mbox{TE}_{20}$ mode in reflection.
\begin{figure*}[htb]
\includegraphics[scale=0.5]{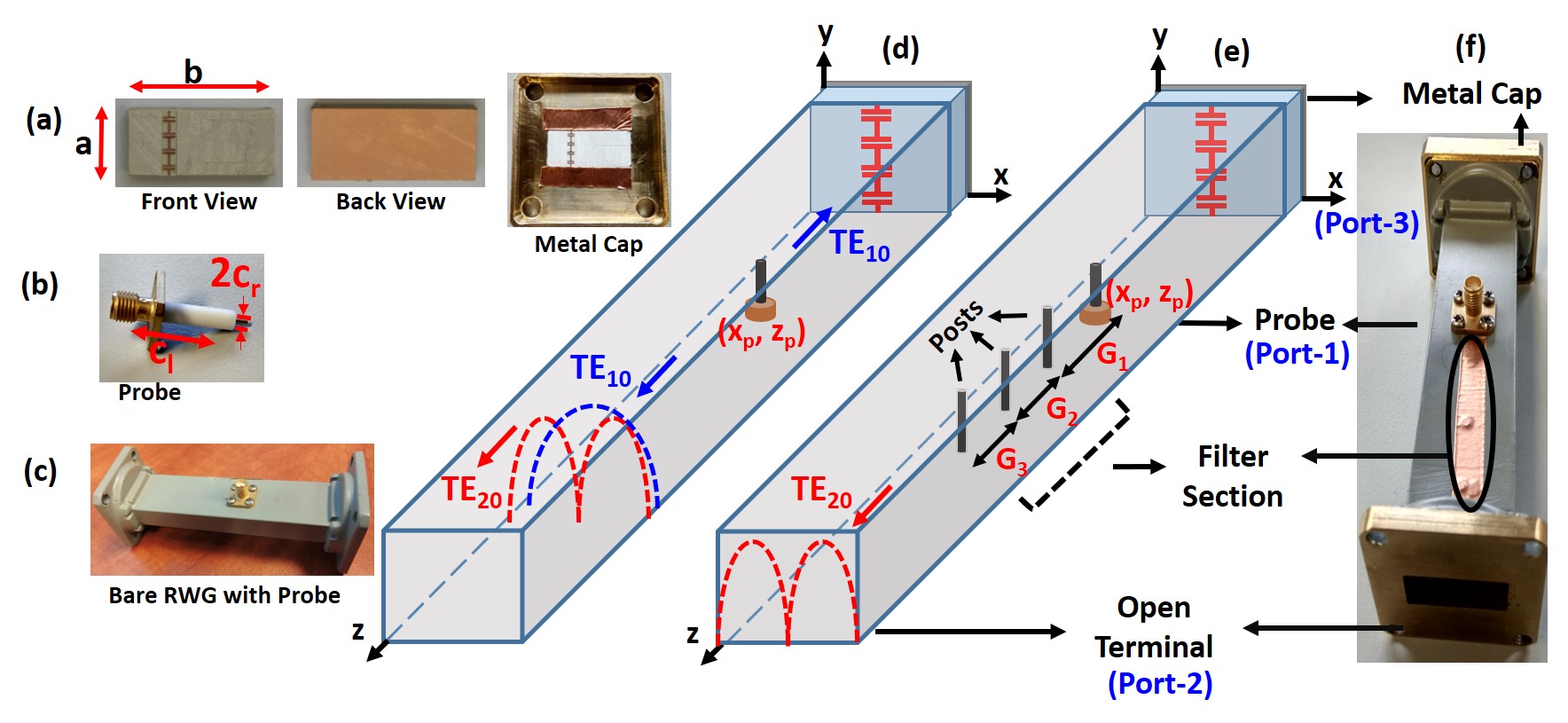}% Here is how to import EPS art
\caption{\label{fig:Fabrication prototype}
PCB MG MC prototype. (a) Front and back view of the fabricated MG PCB along with the metal cap holder used for attaching it to the RWG [port 3 in (e)-(f)].
(b) 4-hole SMA female panel mount connector having a center conductor of radius $C_r$ = 0.635 mm and length $C_l=5.2$ mm (see also Appendix \ref{Appendix-1}).
(c) Host metallic RWG with the SMA connector mounted on the top facet, used to inject the $\mbox{TE}_{10}$ mode into the system.
(d) Schematic of the initial RWG MC design [$(x_p,z_p)$ =$(0.534\lambda,2.419\lambda)$], illustrating the mode interference issue at the output [port 2 in (e)-(f)].
(e) Schematic of the ultimate RWG MC design, including the three metallic posts (filter section) introduced to break the probe radiation symmetry, guaranteeing mode purity  at the output (port 2) and proper evaluation of the conversion efficiency.
(f) Photograph of the fabricated MG-based RWG MC prototype used for experimental validation.
}
\end{figure*}

Having all the design specification in hand from our semianalytical scheme (Sections \ref{sec:Theory} and \ref{subsec:MG simulation}), we now move to fabricate the prototype design.
The PCB MG [Fig. \ref{fig:Fabrication prototype}(a)] is fabricated using a LPKF PCB prototyping machine \textit{Protomat S103} in the Technion.
A suitable hollow RWG (WR-90) made of brass  metal is procured from \textit{IMC Microwave industries Ltd., Petach Tikva, Israel} [Fig. \ref{fig:Fabrication prototype}(c)], and the PCB MG is attached to one of its ports using a metallic fixture produced in our in-house mechanical workshop, yielding the configuration depicted in Fig. \ref{fig:wide1}(a).

When designing the experiment, though, an additional aspect should be taken into consideration. As can be seen in Fig. \ref{fig:E-field plots}, the incident $\mbox{TE}_{10}$ and the converted $\mbox{TE}_{20}$ fields interfere at the input $z=L$, making it hard to quantify the conversion efficiency of the MC when excited with a standard coax-to-waveguide transition from this port.
Therefore, to enable separate measurement of the MG-scattered fields, as well as proper evaluation of the power coupled to each of the modes, we chose to follow a different excitation scheme, including an installed probe and filter sections, as shall be detailed in the following. %the coupled $\mbox{TE}_{20}$ from the exciting $\mbox{TE}_{10}$, we followed a different approach, by incorporating an exciting probe and a filter section into the MC design will be discussed in subsequent paragraphs.

Specifically, to excite the host dual-mode RWG with the fundamental ($\mbox{TE}_{10}$) mode, a 4-hole sub-miniature version A (SMA) panel mount connector (Pasternack, USA, model PE4099), having a center conductor of length $C_l = 5.2$ mm and radius $C_r$ = 0.635 mm [Fig. \ref{fig:Fabrication prototype}(b)], is used as a probe.
Its center conductor penetrates the RWG via a hole drilled on the broader side of the latter [Fig. \ref{fig:Fabrication prototype}(c)].
The probe is installed exactly at the center ($x_p$= a/2) of the RWG's lateral cross-section to achieve maximum power transfer to the $\mbox{TE}_{10}$ mode \cite{balanis2012advanced}; its longitudinal location $z_p=2.419 \lambda$ is chosen with the aid of full-wave simulations, maintaining a sufficient distance from the MG PCB as in Fig. \ref{fig:Fabrication prototype}(d) to reduce undesired near-field coupling (see also Appendix \ref{Appendix-1}) .

This modified RWG assembly, excited by the mounted SMA probe and terminated with the MG PCB, should serve our purpose of experimentally validating the theoretical design.
However, as illustrated in Fig. \ref{fig:Fabrication prototype}(d), even if the MG performs perfect mode conversion, the fields at the output (open) terminal would feature a mixed composition. Specifically, they would be a superposition of the $\mbox{TE}_{10}$ mode directly back-radiated from the probe without interacting with the MG, and the $\mbox{TE}_{20}$ fields converted by the PCB MG termination.
This mode impurity would not allow convenient quantification of the MG performance in the lab, and would also be undesirable application-wise (i.e., if one would wish to consider this device as a $\mbox{TE}_{20}$ generator, engineered based on the proposed MG MC).
To overcome this issue, the probe should be made to excite the configuration asymmetrically, i.e., illuminating only one end of the RWG.
To break this inherent probe-radiation-pattern symmetry, we decided to add a filter section after the probe in order to reflect the probe-excited $\mbox{TE}_{10}$ propagating towards the open end of the RWG.
By restricting the filter structure to the lateral center of the RWG cross-section, where the $\mbox{TE}_{20}$ mode profile has a null, we can ensure that these target fields generated by the MG MC remain unaffected, reaching undisturbed to the output port.

More specifically, following the traditional in-line band-stop filter technique in \cite{matthaei1980microwave,maaskant2016teaching}, we deployed a filter section inside the RWG, at a distance of $G_1$ from the probe, comprising of three successive metallic posts with gaps $G_2$ and $G_3$, as depicted in Fig. \ref{fig:Fabrication prototype}(e).
In general, the distances $G_1$ ,$G_2$ and $G_3$ should match odd multiples of $\lambda_g/4$ to work as a band-stop filter \cite{matthaei1980microwave} , where $\lambda_g$ is the effective wavelength of the $\mathrm{TE}_{10}$ mode to be filtered, given by $\lambda_g=2\pi/\beta_{1,1} = 24.25$ mm; cascading additional posts increases the filter rejection level. These design rules assume multiple reflections of the $\mathrm{TE}_{10}$ modes between infinitesimally-thin PEC wires, i.e. they apply as long as the post radius is much smaller than the effective wavelength $r_0\ll \lambda_g$ and the reactive mutual coupling between the posts is negligible.
Subsequently, we utilize three aluminium posts of radius $r_0=1$ mm, separated by distances of the order of 3$\lambda_g$/4 = $18.188$ mm, a value that was found to provide a reasonable trade-off between the decay length of evanescent modes and the overall filter section length.
After optimizing (Ansys HFSS) the inter-post distances to accommodate the residual near-field coupling, an acceptable $\mbox{TE}_{10}$ rejection level of $-19\,\mathrm{dB}$ was achieved at operating frequency ($f=14$ GHz).

The final specifications of the prototype [Fig. \ref{fig:Fabrication prototype}(e)], including the MG MC, the exciting probe, and the filter section are listed in Table \ref{tab:Design specifications for experimental validation}.
Overall, it functions as a $\mathrm{TE}_{20}$ mode generator, using the MG to convert the $\mathrm{TE}_{10}$ power injected by the probe and reflected by the filter to a pure higher-order mode at the output terminal.

%% Table
\begin{table}
	\caption{\label{tab:Design specifications for experimental validation}
		Design specifications of the RWG $\mbox{TE}_{10}$ - $\mbox{TE}_{20}$ MC for experimental validation (All values are given in mm)}
	\begin{ruledtabular}
		\begin{tabular}{|c|c|c|c|c|c|c|}
			\multicolumn{7}{|c|}{\textbf{MC}}\\\hline
			$a$ &$b$ &$c$ &$x_0$ &$h$ &$W$ &$l$\\\hline
			22.86 & 10.16 & 120 & 6.31 & 2.54 & 1.9 & 2.54 \\
		\end{tabular}
		
		\begin{tabular}{|c|c|c|c|}
			\multicolumn{4}{|c|}{\textbf{Probe}}\\\hline
			$x_p$&$z_p$&$c_l$&$c_r$\\\hline
			11.43&51.831&5.2&0.635\\
		\end{tabular}
		
		\begin{tabular}{|c|c|c|c|}
			\multicolumn{4}{|c|}{\textbf{Filter}}\\\hline
			$G_1$ & $G_2$ & $G_3$ & $r_0$ \\\hline
			15 & 16.5 & 15 & 1 \\
		\end{tabular}
	\end{ruledtabular}
\end{table}

\begin{figure}[htb]
\includegraphics[scale=0.55]{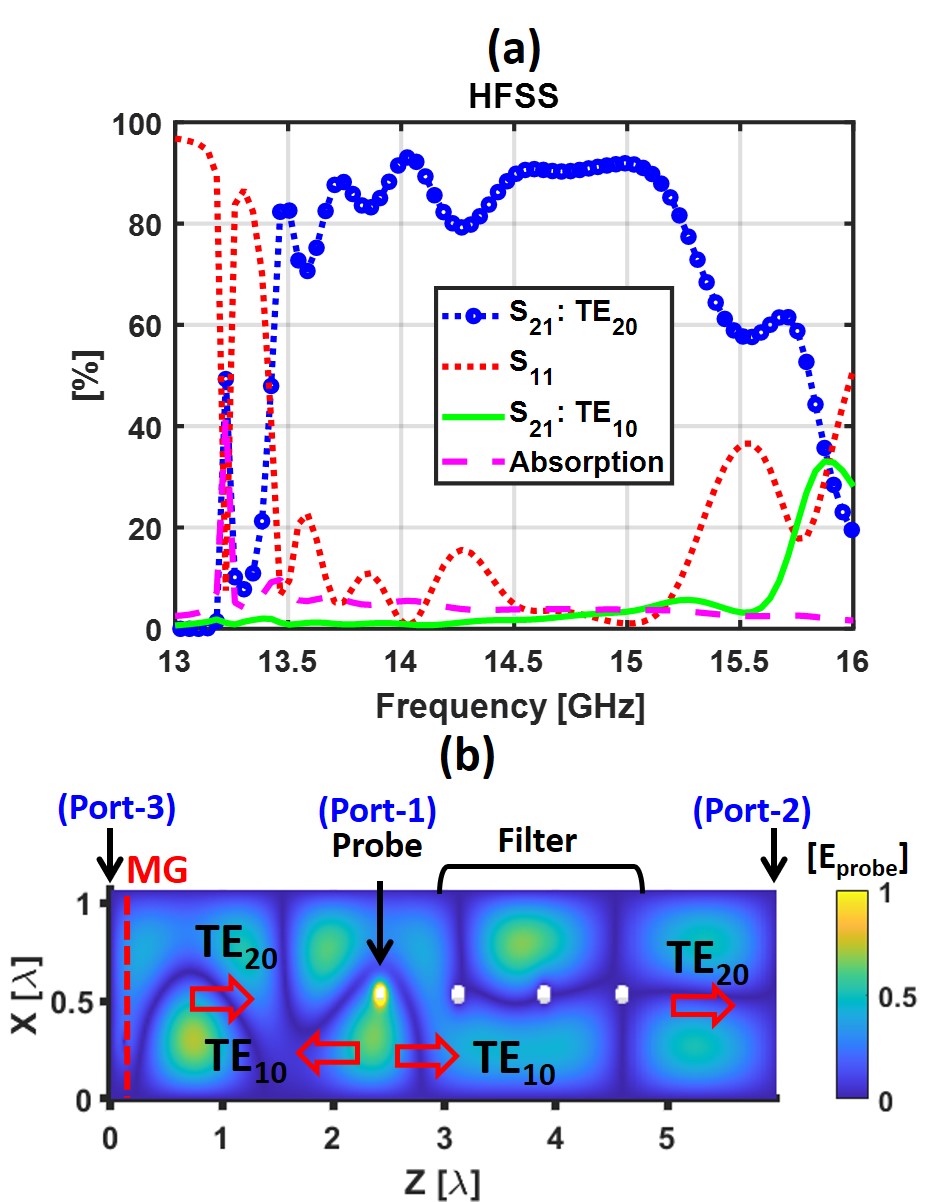}
\caption{\label{fig:frequency response}
Full-wave simulation results for the complete RWG $\mbox{TE}_{10}$-$\mbox{TE}_{20}$ MC configuration depicted in Fig. \ref{fig:Fabrication prototype}(e).
(a) Scattering parameters.
(b) Electric field distribution $\Re[{E_y^\mathrm{tot}(x,z)}]$ at the operating frequency ($f = 14$ GHz).
}
\end{figure}

The finalized design is subsequently defined and simulated in HFSS, placing a wave-port at the output terminal [port 2 in Fig. \ref{fig:Fabrication prototype}(e)] of the RWG, to inspect its mode coupling performance prior to fabrication (see Appendix \ref{Appendix-1} for details on the probe modelling in full-wave solver).
The fraction of input (probe) power coupled to the various modes at the input (port 1) and output (port 2) as recorded in the full-wave solver is presented in Fig. \ref{fig:frequency response}(a) as a function of frequency.
The considered frequency range $[13\,\mathrm{GHz},16\,\mathrm{GHz}]$ allows examination of the MC response within a significant range in the dual-mode operation regime, with the cutoff frequencies of the $\mathrm{TE}_{20}$ and $\mathrm{TE}_{30}$ being $13.11$ GHz and $19.68$ GHz, respectively.

The plot shows that an overall $\mathrm{TE}_\mathrm{20}$ coupling efficiency (from probe to output terminal) of above $\approx 80\%$ is maintained across the range $13.6\mbox{GHz} - 15.2 \mbox{GHz}$ (solid blue line), reaching a peak value of $93.05\%$ at the operating frequency ($f_0=14$ GHz).
The performance is primarily limited by reflections at the probe ($S_{11}$, dotted red), and losses (absorption, dash magenta), with the unconverted $\mbox{TE}_{10}$ ($S_{21}$: $\mbox{TE}_{10}$, solid green) power reaching the output port  restricted to very low values.
The prototype performance as a $\mathrm{TE}_{20}$ generator is further demonstrated by the electric field snapshot presented in Fig. \ref{fig:frequency response}(b), where the characteristic antisymmetric mode profile is clearly observed at the output (port 2).

Finally, the numerically verified configuration is fabricated and assembled to form the prototype device shown in \ref{fig:Fabrication prototype}(f), subsequently used for experimental validation of the proposed concept. This validation is carried out in two phases: first (Section \ref{subsubsec:S-parameters}), by examining the scattering matrix (S-parameters) of the device under test (DUT), and second (Section \ref{subsubsec:Radiation pattern}), by studying its far-field radiation characteristics.

\begin{figure*}[htb]
\includegraphics[scale=0.4]{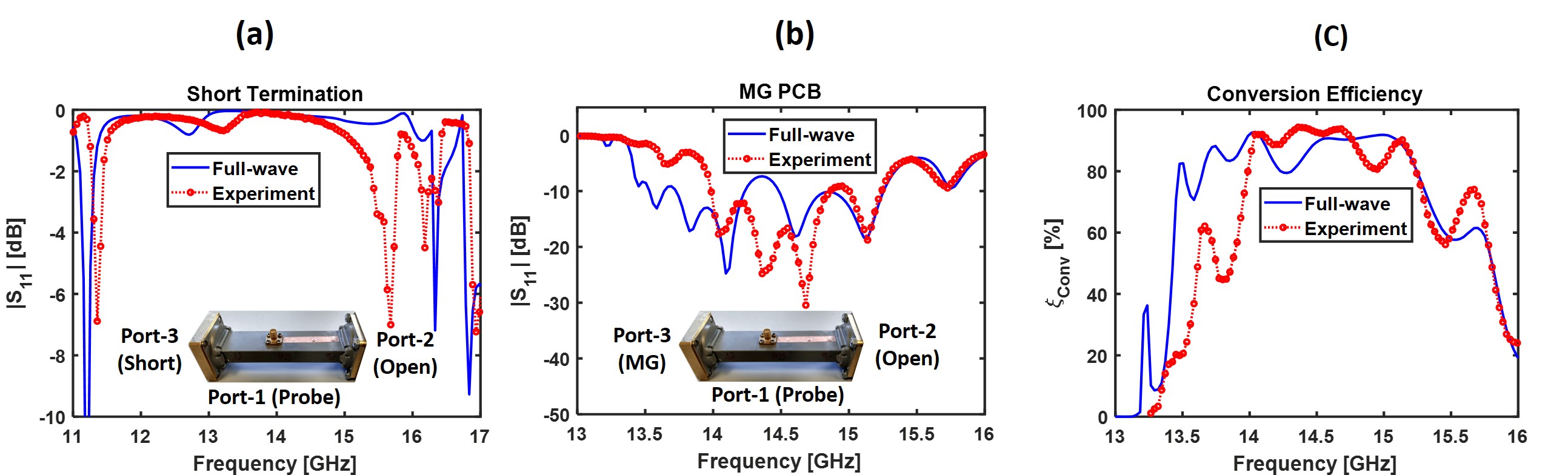}% Here is how to import EPS art
\caption{\label{fig:S11}
Prototype characterization via $S$-parameter measurements (Section \ref{subsubsec:S-parameters}), comparing data extracted from measurements (red dashed line with circle markers) with full-wave simulation results (blue solid line).
Reflection coefficient at the input $S_{11}$ for
(a) the DUT with the MG replaced by a short termination at port 3 (reference) and
(b) the prototype MG-based RWG MC as DUT; insets illustrate the measured configurations.
(c) Conversion efficiency $\xi_{\mathrm{TE}_{20}}^{\mathrm{MC}}$ of the prototype MC.
}
\end{figure*}
\subsubsection{S-parameters}\label{subsubsec:S-parameters}
Although the DUT effectively features two ports (input probe and RWG output terminal), direct measurements of the power coupled to port 2 may be complicated using standard microwave apparatus.
This is because the conventional coax-to-WG transitions, commonly used to connect a vector network analyzer to the RWG, are not designed to incouple well $\mathrm{TE}_{20}$ modes, which are expected to be the main contributors to the power at the output.
Therefore, when characterizing the prototype from a scattering matrix perspective, we consider only the reflection coefficient as observed from the input port, $S_{11}$, measured with a calibrated Keysight P9374A Streamline USB Vector Network Analyzer (VNA), and resort to an indirect scheme to assess the mode conversion efficiency (a complementary, more direct, assessment using far-field measurements will be presented in Section \ref{subsubsec:Radiation pattern}).

As a reference, we first measure the reflection coefficient at the input when the MG is removed and a smooth metallic cap is used to terminate port 3, acting as a short.
Since no conversion to the higher-order $\mbox{TE}_{20}$ mode should occur in this case, we expect most of the power generated by the probe, injected into the RWG  in the form of a $\mbox{TE}_{10}$ mode, to be reflected back to the input port, due to the terminating short on one side and the three-post filter on the other.
Indeed, as can be seen in Fig. \ref{fig:S11}(a), the measured reflection coefficient (dashed red with circle markers) remains very close to unity between $\approx$ 11.5 GHz to 15.1 GHz, indicating the band-stop filter effectiveness in this spectral range.
These results agree quite well with the full-wave simulated $|S_{11}|$ (solid blue), obtained with the finite short-terminated prototype model embedded in a surrounding bound box (radiation boundaries) separated by $\approx2\lambda$ from each of the DUT's facets, and the output terminal left open as in the experiment.
A small blue-shift in the lower frequency range of the figure can be observed in the experimental results, accompanied by some localized discrepancies in notch positions above 15 GHz.
These can be attributed to difficulties in accurate modelling of the probe configuration in Ansys HFSS, possible alignment errors associated with the three-post filter and probe positions, and residual air gaps that may have formed during the assembly process, which might affect resonant phenomena in the system.
Nonetheless, despite minor deviations within the region of interest, both simulated and measured results clearly indicate that no significant $\mbox{TE}_{10}$-$\mbox{TE}_{20}$ mode conversion takes place at the operation frequency in the absence of the MG, and most of the input power is coupled back to the source.

Next, we assemble the prototype in its functional mode-conversion configuration, attaching the PCB MG to port 3 [Fig. \ref{fig:Fabrication prototype}(e) and (f)]. The measured reflection coefficient at the input $|S_{11}|$, shown in Fig. \ref{fig:S11}(b) in the red dashed curve, indicate that in contrast to the case of a short-terminated device, when the designed MG is in place, most of the probe power in the operating frequency is efficiently injected into the DUT. This result is an indication that a major fraction of the power undergoes mode conversion that allows it to pass through the filter without bouncing back into the input port.
In particular, for frequencies between 14 GHz to 15.1 GHz, the probe is well matched and most of the input power is launched into the DUT; this $\mathrm{TE}_{10}$ injection efficiency can be evaluated as $\xi_{\mathrm{in}}^{\mathrm{MC}}=1-|S_{11}|^2$.
As in Fig. \ref{fig:S11}(a), a small blue-shift in the measured results is again noticeable when compared to the full-wave simulated data (solid blue); the observed discrepancies, which are associated with the same factors mentioned in the previous paragraph, have a small effect on the performance near the operating frequency, and do not alter the main conclusions drawn from the figures.

Nevertheless, it is not guaranteed that all of the accepted power will be coupled to the $\mbox{TE}_{20}$ mode and leave the DUT through the output terminal.
This is due to the suboptimal (though very high) conversion efficiency of the MG and the finite rejection level of the filter section, allowing, in principle, some of the input power to propagate to port 2 as $\mathrm{TE}_{10}$ fields.
In addition, some of the power can be dissipated in the realistic (lossy) MG substrate and conductors.
As discussed in the beginning of this subsection, since we do not have full access to the output port fields using standard S-parameter measurements, this modal impurity and absorption loss cannot be directly measured (this will be addressed in Section \ref{subsubsec:Radiation pattern}); however, we can estimate them from the simulated results of Fig. \ref{fig:frequency response}(a) as $\xi_{\mathrm{TE}_{10}}^{\mathrm{MC}}$ and $\xi_{\mathrm{loss}}^{\mathrm{MC}}$, respectively.

Correspondingly, we use these measured and simulated parameters to assess the overall conversion efficiency of the DUT, as deduced from the experimentally obtained S-parameters, via
\begin{equation}\label{eq:Efficiency_S-parameters}
\xi_{\mathrm{TE}_{20}}^{\mathrm{MC}}(f)
=\xi_{\mathrm{in}}^{\mathrm{MC}}(f)
-[\xi_{\mathrm{TE}_{10}}^{\mathrm{MC}}(f)
+\xi_{\mathrm{loss}}^{\mathrm{MC}}(f)]
\end{equation}
and plot the result in Fig. \ref{fig:S11}(c), alongside the fraction of power coupled to the $\mathrm{TE}_{20}$ mode in the DUT as recorded by full-wave simulations [Fig. \ref{fig:frequency response}(a)].
The comparison reveals again the previously discussed slight blue-shift at the lower frequency range, with reasonable agreement between experimental findings and numerical predictions for $f\geq 14$ GHz.
In particular, the results verify that the fabricated prototype effectively converts the injected $\mathrm{TE}_{10}$ mode into the desired higher-order $\mathrm{TE}_{20}$ mode, with peak conversion efficiencies of above $90\%$ around the designated operating frequency, and a fractional bandwidth of $\approx 9\%$ (14 GHz - 15.3 GHz) in which $\xi_{\mathrm{TE}_{20}}^{\mathrm{MC}}\geq 80\%$.

%%%
%%%%%%%%%%%%%%%%%%%%%%%%%%%%%%%%%%%%%%%%%%%%%%%%%%%%%%%%%%%%%%%%%%%%%%%%%%%%%%
\subsubsection{Radiation pattern}\label{subsubsec:Radiation pattern}
\begin{figure}[htb]
\includegraphics[scale=0.26]{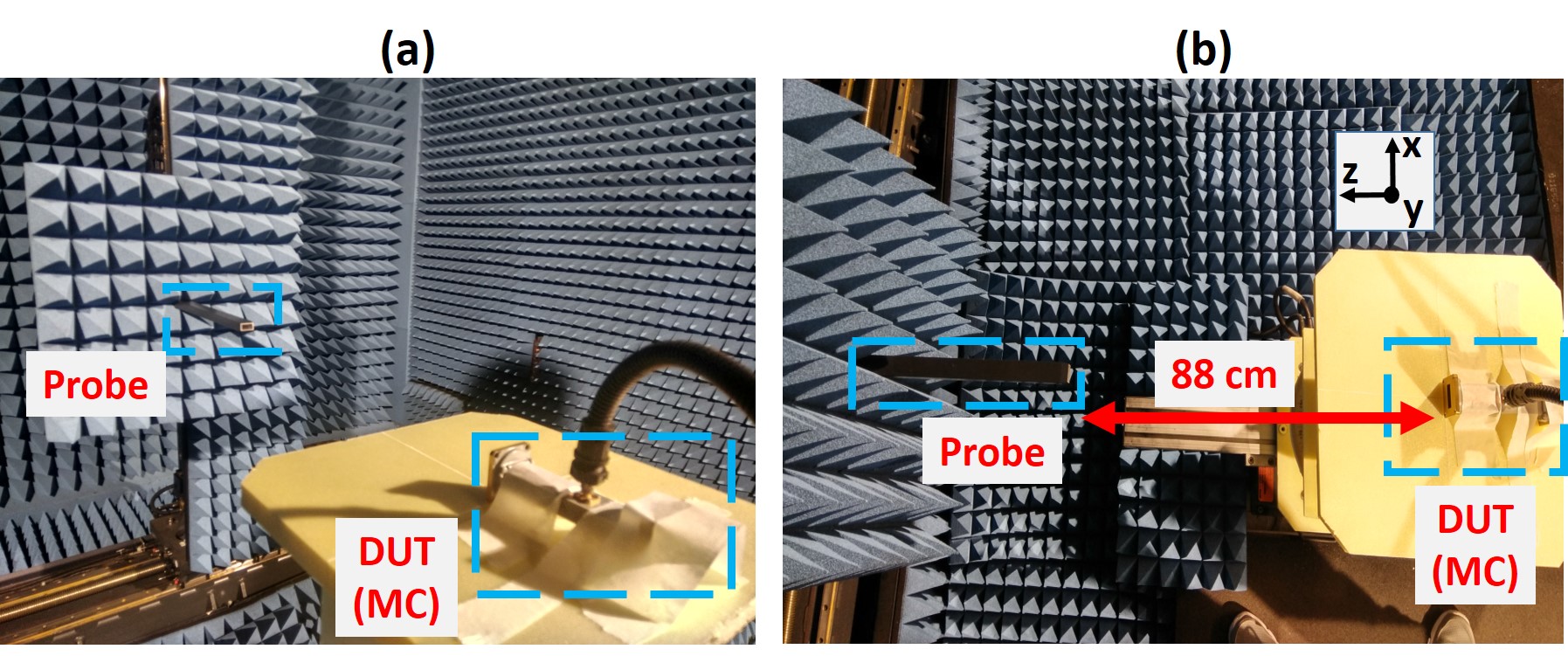}% Here is how to import EPS art
\caption{\label{fig:Characterization}
Experimental setup for radiation pattern measurements (Section \ref{subsubsec:Radiation pattern}). Diagonal (a) and top (b) views from the anechoic chamber, showing the near-field probe and DUT.}
\end{figure}

To corroborate these results, we turn to yet another experimental technique, providing us access to the properties of the fields outcoupled from the DUT at port 2, featuring per design the converted $\mathrm{TE}_{20}$ modes.
More specifically, we measure the radiation characteristics of the prototype in an anechoic chamber at the Technion, equipped with a near-field measurement system (\textit{MVG/Orbit-FR}).
In this sense, the DUT acts as an aperture antenna \cite{balanis2012advanced}, fed by the input probe (port 1) and radiating from the aperture (open output terminal) at the end of the RWG (port 2). As depicted in Fig. \ref{fig:Characterization}, the DUT is placed over a foam stand in front of the system's open waveguide (OWG) near-field probe, at a distance of 88 cm $(\approx41\lambda)$.
The data acquisition module is then programmed to scan the probe along the y-axis, covering the range from -150 cm to 150 cm, while rotating the transmitting DUT in angular steps of $1.4^o$ from $-150^o$ to $150^o$.
Finally, the post-processing made in the system's MiDAS software package utilizes the equivalence principle \cite{balanis2012advanced} to produce the far-field radiation pattern of the DUT from the fields recorded on the measurement surface.

We repeat the same measurement twice. First, with the prototype $\mathrm{TE}_{20}$ mode generator presented in Fig. \ref{fig:Fabrication prototype}(e) and (f) and characterized in Section \ref{subsubsec:S-parameters}, excited by a probe sandwiched between the three-post filter section and the PCB MG MC proposed herein.
Second, with a reference standard (unmodified) WR-90 RWG \emph{identical} to the one used to build the prototype, excited by a commercial coax-to-WG transition from one of its ports.
In the other (output) port of this bare RWG, the fields are expected to be in the form of a pure $\mathrm{TE}_{10}$ mode.
The radiation corresponding to these aperture fields would be used as a reference to the total input power of the near-field measurement system, enabling quantification of the prototype (MC) conversion efficiency.

\begin{figure*}[htb]
\includegraphics[scale=0.42]{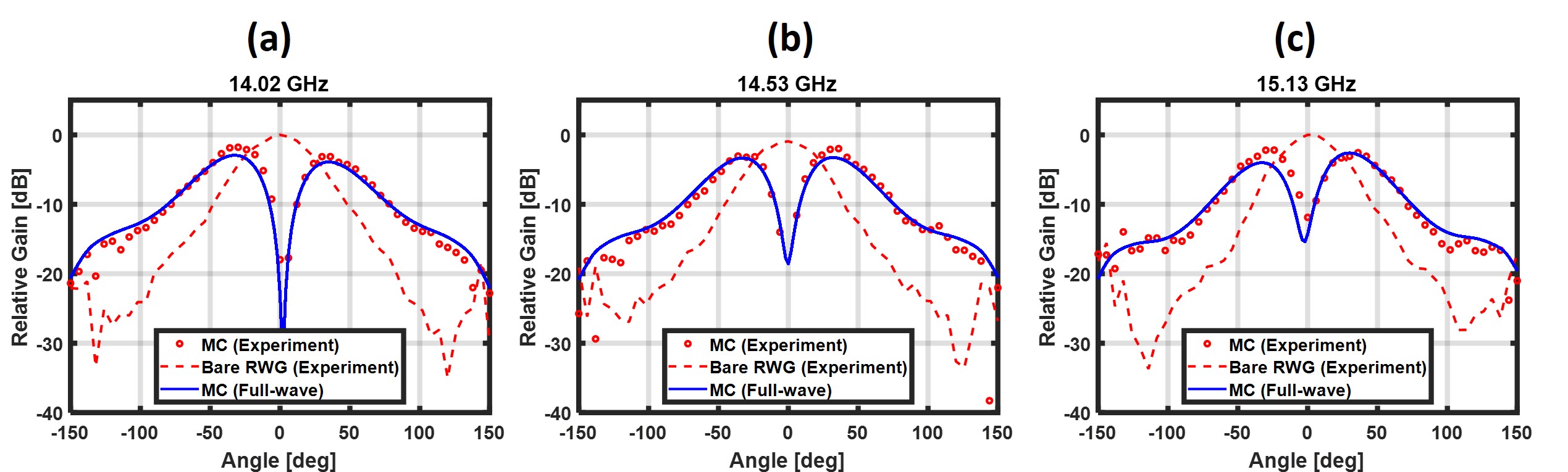}% Here is how to import EPS art
\caption{\label{fig:Radiationpattern}
Experimentally measured gain patterns along the $\widehat{xz}$ plane of the fabricated RWG $\mbox{TE}_{10}$ - $\mbox{TE}_{20}$ MC (red circles) and the reference bare $\mbox{TE}_{10}$ RWG (red dashed line), compared with full-wave simulation results (solid blue line) at three different frequency points:
(a) $f = 14.02$ GHz
(b) $f = 14.53$ GHz, and
(c) $f = 15.13$ GHz. All gain values (in all plots) are presented relative to the peak gain of the reference bare RWG at $14.02$ GHz (a), taken as a reference, set at $0$ dB.
%In these plots, the measured peak gain of the reference bare RWG at $14.02$ GHz is considered as $0$ dB level (reference) and the values in all the graphs are taken relative to this value.
}
\end{figure*}
\begin{figure}[htb]
\includegraphics[scale=0.5]{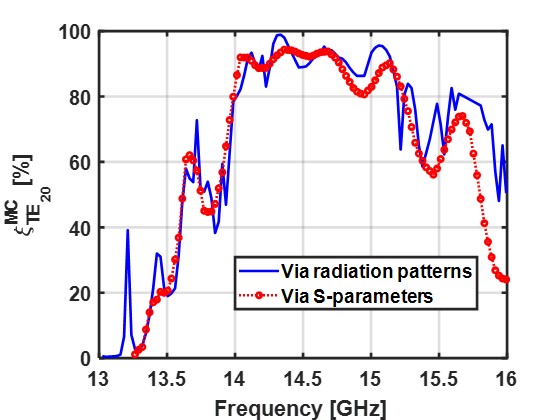}% Here is how to import EPS art
\caption{\label{fig:Efficiency}
Experimentally evaluated conversion efficiency $\xi_{\mathrm{TE}_{20}}^{\mathrm{MC}}$ as assessed from the measured radiation patterns of Fig. \ref{fig:Radiationpattern} as per Eq. \eqref{eq:Efficiency} (solid blue line) and from the measured S-parameters of Fig. \ref{fig:S11} as per Eq. \eqref{eq:Efficiency_S-parameters} (dotted red line with circular markers).
}
\end{figure}
We begin our analysis by considering the DUT radiation patterns at representative frequencies within the operation band.
Correspondingly, Fig. \ref{fig:Radiationpattern} presents the the measured and full-wave (CST microwave studio) simulated radiation patterns of the prototype MC and the reference bare RWG taken along the $\widehat{xz}$ plane at frequencies 14.02 GHz, 14.53 GHz, and 15.13 GHz
\footnote{It is important to include the flange at the output (port 2) of the RWG [Fig. \ref{fig:Fabrication prototype}(e)] in full-wave simulations ; otherwise, increased backscattering effects will be recorded, inconsistent with the measured scenario. Consequently, we have used the following dimensions for the flange, extracted from the fabricated prototype, in our simulations: the width of the metal rim along $x$ and $y$ are taken as $9.1$ mm and $15.45$ mm, respectively, while the thickness along the $z$-axis is kept as $1.27$ mm}.
These plots lead to two important observations.
First, the high mode purity at the output of the devised prototype (red circles) is clearly distinguishable, featuring the signature $\mbox{TE}_{20}$ radiation profile containing a null at broadside and two nearly-symmetric lobes.
The contrast with the classical $\mbox{TE}_{10}$ aperture field radiation patterns \cite{balanis2012advanced} produced by the reference RWG (dashed red) further emphasizes the successful mode conversion taking place within the MG-based device.
Second, the excellent agreement between the radiation patterns obtained in the experiment and the ones recorded in full-wave simulations provides additional evidence for the mode purity of the fields at the fabricated device's output.
The fidelity of the simulated results as reflected in this result implies that the mode composition of the prototype's aperture fields should match the one predicted by CST, which, according to Fig. \ref{fig:frequency response}, indicate a highly-efficient mode conversion process [$\xi_{\mathrm{TE}_{20}}^{\mathrm{MC}}(f_0)\geq 90\%$].

Next, we proceed to obtaining a quantitative assessment of the conversion efficiency $\xi_{\mathrm{TE}_{20}}^{\mathrm{MC}}$, independent of the one presented in Section \ref{subsubsec:S-parameters} based on S-parameter measurements.
To this end, we compare the total power radiated from port 2 of the prototype $P_\mathrm{rad}^\mathrm{MC}$, which practically features only $\mathrm{TE}_{20}$ fields (Figs. \ref{fig:frequency response} and \ref{fig:Radiationpattern}), to the total power injected into the device by the near-field measurement system, $P_{\mathrm{in}}$.
The former can be readily calculated by integrating the measured 3D radiation pattern of the MC prototype $S_\mathrm{rad}^{\mathrm{MC}}(\theta,\phi)$ (red circles in Fig. \ref{fig:Radiationpattern}), reading $P^{\mathrm{MC}}_{\mathrm{rad}}=\int_0^{2\pi}\int_0^{\pi}S_\mathrm{rad}^{\mathrm{MC}}(\theta,\phi)\sin(\theta)d\theta d\phi$.
To evaluate $P_{\mathrm{in}}$, we extract the total power radiated by the bare (reference) RWG $P_\mathrm{rad}^\mathrm{bare}$ from the measured patterns $S_\mathrm{rad}^{\mathrm{bare}}(\theta,\phi)$ (dashed red curves in Fig. \ref{fig:Radiationpattern}), again utilizing $P^{\mathrm{bare}}_{\mathrm{rad}}=\int_0^{2\pi}\int_0^{\pi}S_\mathrm{rad}^{\mathrm{bare}}(\theta,\phi)\sin(\theta)d\theta d\phi$. Considering the reflection coefficient $S_{11}^{\mathrm{bare}}$ of the bare RWG (measured separately) and assuming negligible absorption in this unmodified short WG section, one can estimate the total power injected into the RWG in the measurement as $P_{\mathrm{in}}=P_\mathrm{rad}^\mathrm{bare}/\left[1-\left|S_{11}^\mathrm{bare}\right|^2\right]$.
Consequently, the conversion efficiency of the prototype MC as a function of frequency can be estimated as
\begin{equation}\label{eq:Efficiency}
\begin{split}
\xi_{\mathrm{TE}_{20}}^{\mathrm{MC}}(f)
%&
= \frac{P_{\mathrm{rad}}^{\mathrm{MC}}(f)}{P_{\mathrm{in}}(f)}
%\\
%&
= \frac
{\big[1-\big|S_{11}^{\mathrm{bare}}(f)\big|^2\big]
P^{\mathrm{MC}}_{\mathrm{rad}}(f)}
{P^{\mathrm{bare}}_{\mathrm{rad}}(f)}
\end{split}
\end{equation}

We utilize Eq. \eqref{eq:Efficiency} to calculate the device's conversion efficiency from the radiation pattern measurements. The estimated $\xi_{\mathrm{TE}_{20}}^{\mathrm{MC}}$ as a function of frequency is presented in Fig. \ref{fig:Efficiency} in solid blue line, alongside the values estimated from the S-parameter measurements in Section \ref{subsubsec:S-parameters} [Fig. \ref{fig:S11}(c) therein]. Despite being obtained using completely different techniques, the evaluations closely follow one another, serving as a robust experimental proof for the applicability of our theoretical synthesis scheme (Section \ref{sec:Theory}) in practical scenarios. As verified using both simulated and measured results, the semianalytically designed PCB MG demonstrates highly-efficient overall mode conversion, even when realistic losses and actual feed non-idealities are present.
%Following the equation Eq. \ref{eq:Efficiency}, yields the measured conversion efficiency plots of fabricated prototype (solid blue line) presented in Fig. \ref{fig:Efficiency} comparing with the values measured using S- parameters (dotted line with red circular markers) in the preceding sub-section as a function of frequency.
%The conversion efficiencies evaluated through both the methods are in close agreement with each other further validates our proposed design methodology.

\section{\label{sec:level1}Conclusion\
}\label{sec:Conclusion}
To conclude, we presented a semianalytical scheme to capitalize on the unique properties of MGs for designing RWG $\mbox{TE}_{10}$ to $\mbox{TE}_{20}$ MCs.
Starting with the analytical formulation, the entire synthesis procedure was laid out in detail, including full-wave verification, prototype fabrication, and experimental characterization validating the proposed concept.
As shown, a single capacitively-loaded strip defined on a metal-backed dielectric substrate terminating the dual-mode RWG is sufficient to convert a given input $\mbox{TE}_{10}$ to $\mbox{TE}_{20}$ mode with extremely high efficiency.
This is achieved by judiciously setting the coordinates and geometry of the scatterer following the analytical model as to eliminate reflected $\mbox{TE}_{10}$ fields while retaining power conservation, leading, ideally, to perfect conversion to the higher-order mode.
Similar to previous work on MGs, the semianalytical method directly yields a realistic fabrication-ready PCB layout, without resorting to extensive full-wave optimization.

To validate the fidelity of the developed design scheme, a suitable prototype MC was conceived and fabricated. Due to the inability of standard coax-to-WG transitions to efficiently incouple and distinguish between fundamental and higher-order modes, a specialized MC system was devised, including an SMA probe to inject $\mbox{TE}_{10}$ power into the RWG and a filter section to prevent mode impurity at the output, effectively forming a $\mbox{TE}_{20}$ mode generator based on the designed PCB MG MC. To quantify the conversion efficiency in view of these challenges, two different characterization methods were used, relying on either scattering parameter (reflection coefficient) measurements at the input or radiation pattern recorded at the output. Notably, both techniques agree very well, indicating that the conversion efficiency of the realized device exceeds $90\%$ at the design frequency $f_0=14$ GHz, mainly limited by conductor and dielectric loss.

In contrast to common techniques used for manipulating modes in RWGs, typically including WG deformations subject to numerical optimization, the proposed device features a simple alternative solution, semianalytically designed in the form of a standard PCB. In addition to the practical benefit that such a solution may yield to applications involving RWG, the presented theoretical and experimental findings point out the great potential of MG-based devices for tackling reliably and efficiently a variety of electromagnetic problems in diverse systems.

\begin{acknowledgments}
This work was supported by the Israel Science Foundation (Grant 1540/18). V. K. Killamsetty gratefully acknowledges the financial support of the Tehcnion and the Israel Council for Higher Education.
The authors would like to thank Yuri Komarovsky of the Communication Laboratory and Kalman Maler of the mechanical workshop at the Technion for fabricating and assembling the proposed prototype structure. The authors also wish to thank Ben-Zion Joselson and Denis Dikarov of the Communication Laboratory for helpful discussions. Lastly, they wish to thank Rogers Corporation for providing the laminates used in work.
\end{acknowledgments}

\appendix

\section{Assessing the fields at the MG strip location}\label{Appendix}
As mentioned in Section \ref{subsec:Distributed Load Impedance}, direct substitution of $(x_{0},h)$ for $(x,z)$ in Eq. \eqref{eq:source and image fields} does not facilitate evaluation of the fields over the MG strip, due to the divergence of the source field term [$E_{1}^{\mathrm{source}}(x,z)$] in the frame of the infinitesimal line source approximation (see also Appendix of \cite{Rabinovich2018AnalyticalReflection}).
This will become clearer if the source fields are rewritten as an infinite summation of Hankel functions. Specifically, we decompose the sine products in the first row of Eq. \eqref{eq:source and image fields} into separate harmonic exponents, 
\begin{equation}\label{eq:AppendixEq-1}
	\begin{split}
		&E_{{1}}^{\mathrm{source}}(x,z) =-\frac{k_{1}\eta_{1}I_{0}}{4a}
		\\
		& \cdot\sum_{n=-\infty}^\infty
		\frac{e^{-j\beta_{n,1}(z-h)}}{\beta_{n,1}} \big[e^{-j\frac{n\pi}{a}(x-x_{0})}-e^{-j\frac{n\pi}{a}(x+x_{0})}\big]
%		\sin\bigg(\frac{n\pi}{a}x\bigg)\sin\bigg(\frac{n\pi}{a}x_0\bigg)
	\end{split}
\end{equation}
and apply the Poisson formula \cite{tretyakov2003analytical, epstein2017unveiling} to obtain
\begin{equation}\label{eq:AppendixEq-2}
\begin{split}
&E_{{1}}^{\mathrm{source}}(x,z) =
-\bigg(\frac{k_{1}\eta_{1}I_{0}}{2}\bigg)\\
&\cdot\sum_{n=-\infty}^\infty
\left\{
\begin{split}
&H_{0}^{(2)}\left[k_{1}\sqrt{((x-x_{0})-2an)^2+(z-h)^2}\right]\\
&-H_{0}^{(2)}\left[k_{1}\sqrt{((x+x_{0})-2an)^2+(z-h)^2}\right]
\end{split}
\right\}
\end{split}
\end{equation}

Clearly, the $n=0$ term in Eq. \eqref{eq:AppendixEq-2} diverges at the source coordinate $(x,z)\rightarrow(x_{0},h)$. Therefore,
%The Hankel function diverges at $(x,z)$ = $(x_{0},h)$, and therefore, 
following \cite{tretyakov2003analytical,Rabinovich2018AnalyticalReflection}, the self induced fields the MG strip generates on its shell should be treated separately from the fields due to multiple reflections at the RWG metallic walls ($x=0$ and $x=a$), which converge well. More specifically, when assessing the self-induced fields we must deviate from the ideal infinitesimal line-source approximation and consider the actual dimensions of the conducting strips, featuring an effective radius of $r_\mathrm{eff}=w/4$ \cite{tretyakov2003analytical}, where $w$ is the strip width [Fig. \ref{fig:wide1}(b)]. Applying this distinction to the expression in Eq. \eqref{eq:AppendixEq-2} yields
\begin{equation}\label{eq:AppendixEq-3}
\begin{split}
&E_{{1}}^{\mathrm{source}}(x\rightarrow x_{0},z \rightarrow h) =-\bigg(\frac{k_{1}\eta_{1}I_{0}}{2}\bigg)
\\
&\cdot
\left\{
\begin{split}
&H_{0}^{(2)}[k_{1}r_{\mathrm{eff}}]+\sum_{\substack{n=-\infty \\ n\neq 0}}^{\infty}
H_{0}^{(2)}[k_{1}|2an|]\\
&-\sum_{n=-\infty}^{\infty}
H_{0}^{(2)}[k_{1}|2x_{0}-2an|]
\end{split}
\right\}\\
\end{split}
\end{equation}
%%
%where the first and second term of infinite Hankel summation in Eq. \eqref{eq:AppendixEq-3} correspond to the fields generated by the MG strip over itself and the fields generated by the image sources due to the presence of metallic mirrors on both sides of the MG strip along the $x$-axis. 

Subsequently, we continue as in \cite{tretyakov2003analytical}, resolving the first term by using the asymptotic approximation of the Hankel function for small
%The first term is the diverging term (for $n\rightarrow 0$) and is treated following the Hankel function approximation for smaller 
arguments ($r_{\mathrm{eff}}=w/4 \ll \lambda $) \cite{abramowitz1970handbook}, while terms in the infinite summation are expanded as per \cite{jeffrey2007table}.
Carefully applying these transformations into Eq. \eqref{eq:source and image fields} for ($x \rightarrow x_{0}$, $z \rightarrow h$) leads to Eq. \eqref{eq:source and image fields regular} in the main text.

\begin{figure}[b]
	\includegraphics[scale=0.55]{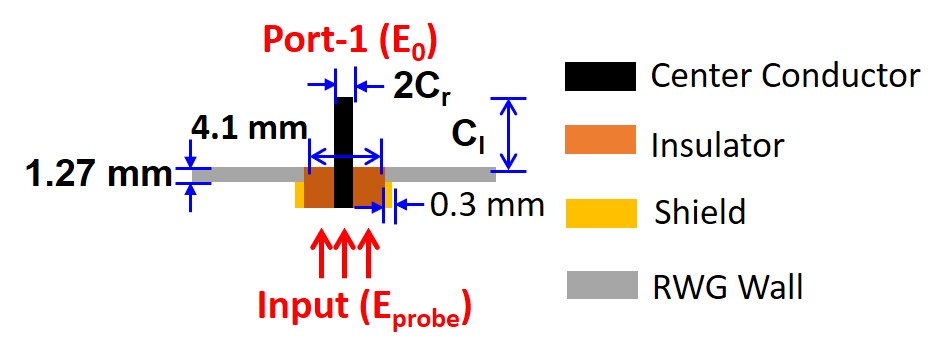}% Here is how to import EPS art
	\caption{\label{fig:modelling probe}
	Model of the exciting probe [Fig. \ref{fig:Fabrication prototype}(b)] for full-wave solvers.}
\end{figure}

\section{Modelling the Exciting Probe in Full-wave Simulations}\label{Appendix-1}
For completeness, we provide herein details regarding the model used to include the exciting SMA probe [Fig. \ref{fig:Fabrication prototype}] in the full-wave simulations (Ansys HFSS and CST Microwave Studio) carried out in Section \ref{sec:results and discussion}.
The probe is defined in full-wave solvers following the schematic presented in Fig. \ref{fig:modelling probe}, showing a cross-section of the azimuthally-symmetric configuration.
In particular, the probe is composed of a center conductor made of copper (cylinder with radius $C_r=0.635$ mm), Polytetrafluoroethylene (PTFE) insulator having a dielectric constant of $\epsilon_r = 2.1$, and a copper shield (dimensions given in Table \ref{tab:Design specifications for experimental validation}).
The insulator shown in the picture of the purchased probe [Fig. \ref{fig:Fabrication prototype}(b)] is cut such that it is aligned with the WG wall when the probe is mounted.
The length of the center conductor $C_l$ entering the RWG is set using full-wave optimization to guarantee impedance matching, applied subsequently to the actual probe using a standard wire cutter.
All other geometrical dimensions are taken from actual measurements of the PE4099 model SMA connector used in the experiment.

%\nocite{*}


\begin{thebibliography}{0}%
\makeatletter
\providecommand \@ifxundefined [1]{%
 \@ifx{#1\undefined}
}%
\providecommand \@ifnum [1]{%
 \ifnum #1\expandafter \@firstoftwo
 \else \expandafter \@secondoftwo
 \fi
}%
\providecommand \@ifx [1]{%
 \ifx #1\expandafter \@firstoftwo
 \else \expandafter \@secondoftwo
 \fi
}%
\providecommand \natexlab [1]{#1}%
\providecommand \enquote  [1]{``#1''}%
\providecommand \bibnamefont  [1]{#1}%
\providecommand \bibfnamefont [1]{#1}%
\providecommand \citenamefont [1]{#1}%
\providecommand \href@noop [0]{\@secondoftwo}%
\providecommand \href [0]{\begingroup \@sanitize@url \@href}%
\providecommand \@href[1]{\@@startlink{#1}\@@href}%
\providecommand \@@href[1]{\endgroup#1\@@endlink}%
\providecommand \@sanitize@url [0]{\catcode `\\12\catcode `\$12\catcode
  `\&12\catcode `\#12\catcode `\^12\catcode `\_12\catcode `\%12\relax}%
\providecommand \@@startlink[1]{}%
\providecommand \@@endlink[0]{}%
\providecommand \url  [0]{\begingroup\@sanitize@url \@url }%
\providecommand \@url [1]{\endgroup\@href {#1}{\urlprefix }}%
\providecommand \urlprefix  [0]{URL }%
\providecommand \Eprint [0]{\href }%
\providecommand \doibase [0]{https://doi.org/}%
\providecommand \selectlanguage [0]{\@gobble}%
\providecommand \bibinfo  [0]{\@secondoftwo}%
\providecommand \bibfield  [0]{\@secondoftwo}%
\providecommand \translation [1]{[#1]}%
\providecommand \BibitemOpen [0]{}%
\providecommand \bibitemStop [0]{}%
\providecommand \bibitemNoStop [0]{.\EOS\space}%
\providecommand \EOS [0]{\spacefactor3000\relax}%
\providecommand \BibitemShut  [1]{\csname bibitem#1\endcsname}%
\let\auto@bib@innerbib\@empty
%</preamble>
\end{thebibliography}%


\begin{thebibliography}{79}%
\makeatletter
\providecommand \@ifxundefined [1]{%
 \@ifx{#1\undefined}
}%
\providecommand \@ifnum [1]{%
 \ifnum #1\expandafter \@firstoftwo
 \else \expandafter \@secondoftwo
 \fi
}%
\providecommand \@ifx [1]{%
 \ifx #1\expandafter \@firstoftwo
 \else \expandafter \@secondoftwo
 \fi
}%
\providecommand \natexlab [1]{#1}%
\providecommand \enquote  [1]{``#1''}%
\providecommand \bibnamefont  [1]{#1}%
\providecommand \bibfnamefont [1]{#1}%
\providecommand \citenamefont [1]{#1}%
\providecommand \href@noop [0]{\@secondoftwo}%
\providecommand \href [0]{\begingroup \@sanitize@url \@href}%
\providecommand \@href[1]{\@@startlink{#1}\@@href}%
\providecommand \@@href[1]{\endgroup#1\@@endlink}%
\providecommand \@sanitize@url [0]{\catcode `\\12\catcode `\$12\catcode
  `\&12\catcode `\#12\catcode `\^12\catcode `\_12\catcode `\%12\relax}%
\providecommand \@@startlink[1]{}%
\providecommand \@@endlink[0]{}%
\providecommand \url  [0]{\begingroup\@sanitize@url \@url }%
\providecommand \@url [1]{\endgroup\@href {#1}{\urlprefix }}%
\providecommand \urlprefix  [0]{URL }%
\providecommand \Eprint [0]{\href }%
\providecommand \doibase [0]{https://doi.org/}%
\providecommand \selectlanguage [0]{\@gobble}%
\providecommand \bibinfo  [0]{\@secondoftwo}%
\providecommand \bibfield  [0]{\@secondoftwo}%
\providecommand \translation [1]{[#1]}%
\providecommand \BibitemOpen [0]{}%
\providecommand \bibitemStop [0]{}%
\providecommand \bibitemNoStop [0]{.\EOS\space}%
\providecommand \EOS [0]{\spacefactor3000\relax}%
\providecommand \BibitemShut  [1]{\csname bibitem#1\endcsname}%
\let\auto@bib@innerbib\@empty
%</preamble>
\bibitem [{\citenamefont {Kuester}\ \emph {et~al.}(2003)\citenamefont
  {Kuester}, \citenamefont {Mohamed}, \citenamefont {Piket-May},\ and\
  \citenamefont {Holloway}}]{kuester2003averaged}%
  \BibitemOpen
  \bibfield  {author} {\bibinfo {author} {\bibfnamefont {E.~F.}\ \bibnamefont
  {Kuester}}, \bibinfo {author} {\bibfnamefont {M.~A.}\ \bibnamefont
  {Mohamed}}, \bibinfo {author} {\bibfnamefont {M.}~\bibnamefont {Piket-May}},\
  and\ \bibinfo {author} {\bibfnamefont {C.~L.}\ \bibnamefont {Holloway}},\
  }\bibfield  {title} {\bibinfo {title} {Averaged transition conditions for
  electromagnetic fields at a metafilm},\ }\href@noop {} {\bibfield  {journal}
  {\bibinfo  {journal} {IEEE Trans. Antennas Propag.}\ }\textbf {\bibinfo
  {volume} {51}},\ \bibinfo {pages} {2641} (\bibinfo {year}
  {2003})}\BibitemShut {NoStop}%
\bibitem [{\citenamefont {Glybovski}\ \emph {et~al.}(2016)\citenamefont
  {Glybovski}, \citenamefont {Tretyakov}, \citenamefont {Belov}, \citenamefont
  {Kivshar},\ and\ \citenamefont {Simovski}}]{glybovski2016metasurfaces}%
  \BibitemOpen
  \bibfield  {author} {\bibinfo {author} {\bibfnamefont {S.~B.}\ \bibnamefont
  {Glybovski}}, \bibinfo {author} {\bibfnamefont {S.~A.}\ \bibnamefont
  {Tretyakov}}, \bibinfo {author} {\bibfnamefont {P.~A.}\ \bibnamefont
  {Belov}}, \bibinfo {author} {\bibfnamefont {Y.~S.}\ \bibnamefont {Kivshar}},\
  and\ \bibinfo {author} {\bibfnamefont {C.~R.}\ \bibnamefont {Simovski}},\
  }\bibfield  {title} {\bibinfo {title} {Metasurfaces: From microwaves to
  visible},\ }\href@noop {} {\bibfield  {journal} {\bibinfo  {journal} {Phys.
  Rep.}\ }\textbf {\bibinfo {volume} {634}},\ \bibinfo {pages} {1} (\bibinfo
  {year} {2016})}\BibitemShut {NoStop}%
\bibitem [{\citenamefont {Asadchy}\ \emph
  {et~al.}(2015{\natexlab{a}})\citenamefont {Asadchy}, \citenamefont {Ra’Di},
  \citenamefont {Vehmas},\ and\ \citenamefont
  {Tretyakov}}]{asadchy2015functional}%
  \BibitemOpen
  \bibfield  {author} {\bibinfo {author} {\bibfnamefont {V.~S.}\ \bibnamefont
  {Asadchy}}, \bibinfo {author} {\bibfnamefont {Y.}~\bibnamefont {Ra’Di}},
  \bibinfo {author} {\bibfnamefont {J.}~\bibnamefont {Vehmas}},\ and\ \bibinfo
  {author} {\bibfnamefont {S.}~\bibnamefont {Tretyakov}},\ }\bibfield  {title}
  {\bibinfo {title} {Functional metamirrors using bianisotropic elements},\
  }\href@noop {} {\bibfield  {journal} {\bibinfo  {journal} {Phys. Rev. Lett.}\
  }\textbf {\bibinfo {volume} {114}},\ \bibinfo {pages} {095503} (\bibinfo
  {year} {2015}{\natexlab{a}})}\BibitemShut {NoStop}%
\bibitem [{\citenamefont {Asadchy}\ \emph {et~al.}(2016)\citenamefont
  {Asadchy}, \citenamefont {Albooyeh}, \citenamefont {Tcvetkova}, \citenamefont
  {D{\'\i}az-Rubio}, \citenamefont {Ra'di},\ and\ \citenamefont
  {Tretyakov}}]{asadchy2016perfect}%
  \BibitemOpen
  \bibfield  {author} {\bibinfo {author} {\bibfnamefont {V.~S.}\ \bibnamefont
  {Asadchy}}, \bibinfo {author} {\bibfnamefont {M.}~\bibnamefont {Albooyeh}},
  \bibinfo {author} {\bibfnamefont {S.~N.}\ \bibnamefont {Tcvetkova}}, \bibinfo
  {author} {\bibfnamefont {A.}~\bibnamefont {D{\'\i}az-Rubio}}, \bibinfo
  {author} {\bibfnamefont {Y.}~\bibnamefont {Ra'di}},\ and\ \bibinfo {author}
  {\bibfnamefont {S.}~\bibnamefont {Tretyakov}},\ }\bibfield  {title} {\bibinfo
  {title} {Perfect control of reflection and refraction using spatially
  dispersive metasurfaces},\ }\href@noop {} {\bibfield  {journal} {\bibinfo
  {journal} {Phys. Rev. B}\ }\textbf {\bibinfo {volume} {94}},\ \bibinfo
  {pages} {075142} (\bibinfo {year} {2016})}\BibitemShut {NoStop}%
\bibitem [{\citenamefont {Estakhri}\ and\ \citenamefont
  {Al{\`u}}(2016)}]{estakhri2016wave}%
  \BibitemOpen
  \bibfield  {author} {\bibinfo {author} {\bibfnamefont {N.~M.}\ \bibnamefont
  {Estakhri}}\ and\ \bibinfo {author} {\bibfnamefont {A.}~\bibnamefont
  {Al{\`u}}},\ }\bibfield  {title} {\bibinfo {title} {Wave-front transformation
  with gradient metasurfaces},\ }\href@noop {} {\bibfield  {journal} {\bibinfo
  {journal} {Phys. Rev. X}\ }\textbf {\bibinfo {volume} {6}},\ \bibinfo {pages}
  {041008} (\bibinfo {year} {2016})}\BibitemShut {NoStop}%
\bibitem [{\citenamefont {Epstein}\ and\ \citenamefont
  {Eleftheriades}(2016{\natexlab{a}})}]{epstein2016synthesis}%
  \BibitemOpen
  \bibfield  {author} {\bibinfo {author} {\bibfnamefont {A.}~\bibnamefont
  {Epstein}}\ and\ \bibinfo {author} {\bibfnamefont {G.~V.}\ \bibnamefont
  {Eleftheriades}},\ }\bibfield  {title} {\bibinfo {title} {Synthesis of
  passive lossless metasurfaces using auxiliary fields for reflectionless beam
  splitting and perfect reflection},\ }\href@noop {} {\bibfield  {journal}
  {\bibinfo  {journal} {Phys. Rev. Lett.}\ }\textbf {\bibinfo {volume} {117}},\
  \bibinfo {pages} {256103} (\bibinfo {year} {2016}{\natexlab{a}})}\BibitemShut
  {NoStop}%
\bibitem [{\citenamefont {Yu}\ \emph {et~al.}(2011)\citenamefont {Yu},
  \citenamefont {Genevet}, \citenamefont {Kats}, \citenamefont {Aieta},
  \citenamefont {Tetienne}, \citenamefont {Capasso},\ and\ \citenamefont
  {Gaburro}}]{yu2011light}%
  \BibitemOpen
  \bibfield  {author} {\bibinfo {author} {\bibfnamefont {N.}~\bibnamefont
  {Yu}}, \bibinfo {author} {\bibfnamefont {P.}~\bibnamefont {Genevet}},
  \bibinfo {author} {\bibfnamefont {M.~A.}\ \bibnamefont {Kats}}, \bibinfo
  {author} {\bibfnamefont {F.}~\bibnamefont {Aieta}}, \bibinfo {author}
  {\bibfnamefont {J.-P.}\ \bibnamefont {Tetienne}}, \bibinfo {author}
  {\bibfnamefont {F.}~\bibnamefont {Capasso}},\ and\ \bibinfo {author}
  {\bibfnamefont {Z.}~\bibnamefont {Gaburro}},\ }\bibfield  {title} {\bibinfo
  {title} {Light propagation with phase discontinuities: generalized laws of
  reflection and refraction},\ }\href@noop {} {\bibfield  {journal} {\bibinfo
  {journal} {Science}\ }\textbf {\bibinfo {volume} {334}},\ \bibinfo {pages}
  {333} (\bibinfo {year} {2011})}\BibitemShut {NoStop}%
\bibitem [{\citenamefont {Pfeiffer}\ and\ \citenamefont
  {Grbic}(2013)}]{pfeiffer2013metamaterial}%
  \BibitemOpen
  \bibfield  {author} {\bibinfo {author} {\bibfnamefont {C.}~\bibnamefont
  {Pfeiffer}}\ and\ \bibinfo {author} {\bibfnamefont {A.}~\bibnamefont
  {Grbic}},\ }\bibfield  {title} {\bibinfo {title} {Metamaterial huygens’
  surfaces: tailoring wave fronts with reflectionless sheets},\ }\href@noop {}
  {\bibfield  {journal} {\bibinfo  {journal} {Phys. Rev. Lett.}\ }\textbf
  {\bibinfo {volume} {110}},\ \bibinfo {pages} {197401} (\bibinfo {year}
  {2013})}\BibitemShut {NoStop}%
\bibitem [{\citenamefont {Monticone}\ \emph {et~al.}(2013)\citenamefont
  {Monticone}, \citenamefont {Estakhri},\ and\ \citenamefont
  {Alu}}]{monticone2013full}%
  \BibitemOpen
  \bibfield  {author} {\bibinfo {author} {\bibfnamefont {F.}~\bibnamefont
  {Monticone}}, \bibinfo {author} {\bibfnamefont {N.~M.}\ \bibnamefont
  {Estakhri}},\ and\ \bibinfo {author} {\bibfnamefont {A.}~\bibnamefont
  {Alu}},\ }\bibfield  {title} {\bibinfo {title} {Full control of nanoscale
  optical transmission with a composite metascreen},\ }\href@noop {} {\bibfield
   {journal} {\bibinfo  {journal} {Phys. Rev. Lett.}\ }\textbf {\bibinfo
  {volume} {110}},\ \bibinfo {pages} {203903} (\bibinfo {year}
  {2013})}\BibitemShut {NoStop}%
\bibitem [{\citenamefont {Selvanayagam}\ and\ \citenamefont
  {Eleftheriades}(2013)}]{selvanayagam2013discontinuous}%
  \BibitemOpen
  \bibfield  {author} {\bibinfo {author} {\bibfnamefont {M.}~\bibnamefont
  {Selvanayagam}}\ and\ \bibinfo {author} {\bibfnamefont {G.~V.}\ \bibnamefont
  {Eleftheriades}},\ }\bibfield  {title} {\bibinfo {title} {Discontinuous
  electromagnetic fields using orthogonal electric and magnetic currents for
  wavefront manipulation},\ }\href@noop {} {\bibfield  {journal} {\bibinfo
  {journal} {Opt. Express}\ }\textbf {\bibinfo {volume} {21}},\ \bibinfo
  {pages} {14409} (\bibinfo {year} {2013})}\BibitemShut {NoStop}%
\bibitem [{\citenamefont {Epstein}\ and\ \citenamefont
  {Eleftheriades}(2016{\natexlab{b}})}]{epstein2016arbitrary}%
  \BibitemOpen
  \bibfield  {author} {\bibinfo {author} {\bibfnamefont {A.}~\bibnamefont
  {Epstein}}\ and\ \bibinfo {author} {\bibfnamefont {G.~V.}\ \bibnamefont
  {Eleftheriades}},\ }\bibfield  {title} {\bibinfo {title} {Arbitrary
  power-conserving field transformations with passive lossless omega-type
  bianisotropic metasurfaces},\ }\href@noop {} {\bibfield  {journal} {\bibinfo
  {journal} {IEEE Trans. Antennas Propag.}\ }\textbf {\bibinfo {volume} {64}},\
  \bibinfo {pages} {3880} (\bibinfo {year} {2016}{\natexlab{b}})}\BibitemShut
  {NoStop}%
\bibitem [{\citenamefont {Pfeiffer}\ \emph {et~al.}(2016)\citenamefont
  {Pfeiffer}, \citenamefont {Zhang}, \citenamefont {Ray}, \citenamefont {Guo},\
  and\ \citenamefont {Grbic}}]{pfeiffer2016polarization}%
  \BibitemOpen
  \bibfield  {author} {\bibinfo {author} {\bibfnamefont {C.}~\bibnamefont
  {Pfeiffer}}, \bibinfo {author} {\bibfnamefont {C.}~\bibnamefont {Zhang}},
  \bibinfo {author} {\bibfnamefont {V.}~\bibnamefont {Ray}}, \bibinfo {author}
  {\bibfnamefont {L.~J.}\ \bibnamefont {Guo}},\ and\ \bibinfo {author}
  {\bibfnamefont {A.}~\bibnamefont {Grbic}},\ }\bibfield  {title} {\bibinfo
  {title} {Polarization rotation with ultra-thin bianisotropic metasurfaces},\
  }\href@noop {} {\bibfield  {journal} {\bibinfo  {journal} {Optica}\ }\textbf
  {\bibinfo {volume} {3}},\ \bibinfo {pages} {427} (\bibinfo {year}
  {2016})}\BibitemShut {NoStop}%
\bibitem [{\citenamefont {Epstein}\ \emph {et~al.}(2016)\citenamefont
  {Epstein}, \citenamefont {Wong},\ and\ \citenamefont
  {Eleftheriades}}]{epstein2016cavity}%
  \BibitemOpen
  \bibfield  {author} {\bibinfo {author} {\bibfnamefont {A.}~\bibnamefont
  {Epstein}}, \bibinfo {author} {\bibfnamefont {J.~P.}\ \bibnamefont {Wong}},\
  and\ \bibinfo {author} {\bibfnamefont {G.~V.}\ \bibnamefont
  {Eleftheriades}},\ }\bibfield  {title} {\bibinfo {title} {Cavity-excited
  huygens’ metasurface antennas for near-unity aperture illumination
  efficiency from arbitrarily large apertures},\ }\href@noop {} {\bibfield
  {journal} {\bibinfo  {journal} {Nat. Commun.}\ }\textbf {\bibinfo {volume}
  {7}},\ \bibinfo {pages} {1} (\bibinfo {year} {2016})}\BibitemShut {NoStop}%
\bibitem [{\citenamefont {Raeker}\ and\ \citenamefont
  {Rudolph}(2016)}]{raeker2016arbitrary}%
  \BibitemOpen
  \bibfield  {author} {\bibinfo {author} {\bibfnamefont {B.~O.}\ \bibnamefont
  {Raeker}}\ and\ \bibinfo {author} {\bibfnamefont {S.~M.}\ \bibnamefont
  {Rudolph}},\ }\bibfield  {title} {\bibinfo {title} {Arbitrary transformation
  of radiation patterns using a spherical impedance metasurface},\ }\href@noop
  {} {\bibfield  {journal} {\bibinfo  {journal} {IEEE Trans. Antennas Propag.}\
  }\textbf {\bibinfo {volume} {64}},\ \bibinfo {pages} {5243} (\bibinfo {year}
  {2016})}\BibitemShut {NoStop}%
\bibitem [{\citenamefont {Minatti}\ \emph {et~al.}(2016)\citenamefont
  {Minatti}, \citenamefont {Caminita}, \citenamefont {Martini}, \citenamefont
  {Sabbadini},\ and\ \citenamefont {Maci}}]{minatti2016synthesis}%
  \BibitemOpen
  \bibfield  {author} {\bibinfo {author} {\bibfnamefont {G.}~\bibnamefont
  {Minatti}}, \bibinfo {author} {\bibfnamefont {F.}~\bibnamefont {Caminita}},
  \bibinfo {author} {\bibfnamefont {E.}~\bibnamefont {Martini}}, \bibinfo
  {author} {\bibfnamefont {M.}~\bibnamefont {Sabbadini}},\ and\ \bibinfo
  {author} {\bibfnamefont {S.}~\bibnamefont {Maci}},\ }\bibfield  {title}
  {\bibinfo {title} {Synthesis of modulated-metasurface antennas with
  amplitude, phase, and polarization control},\ }\href@noop {} {\bibfield
  {journal} {\bibinfo  {journal} {IEEE Trans. Antennas Propag.}\ }\textbf
  {\bibinfo {volume} {64}},\ \bibinfo {pages} {3907} (\bibinfo {year}
  {2016})}\BibitemShut {NoStop}%
\bibitem [{\citenamefont {Wakatsuchi}\ \emph {et~al.}(2013)\citenamefont
  {Wakatsuchi}, \citenamefont {Kim}, \citenamefont {Rushton},\ and\
  \citenamefont {Sievenpiper}}]{wakatsuchi2013waveform}%
  \BibitemOpen
  \bibfield  {author} {\bibinfo {author} {\bibfnamefont {H.}~\bibnamefont
  {Wakatsuchi}}, \bibinfo {author} {\bibfnamefont {S.}~\bibnamefont {Kim}},
  \bibinfo {author} {\bibfnamefont {J.~J.}\ \bibnamefont {Rushton}},\ and\
  \bibinfo {author} {\bibfnamefont {D.~F.}\ \bibnamefont {Sievenpiper}},\
  }\bibfield  {title} {\bibinfo {title} {Waveform-dependent absorbing
  metasurfaces},\ }\href@noop {} {\bibfield  {journal} {\bibinfo  {journal}
  {Phys. Rev. Lett.}\ }\textbf {\bibinfo {volume} {111}},\ \bibinfo {pages}
  {245501} (\bibinfo {year} {2013})}\BibitemShut {NoStop}%
\bibitem [{\citenamefont {Ra’Di}\ \emph {et~al.}(2015)\citenamefont
  {Ra’Di}, \citenamefont {Simovski},\ and\ \citenamefont
  {Tretyakov}}]{ra2015thin}%
  \BibitemOpen
  \bibfield  {author} {\bibinfo {author} {\bibfnamefont {Y.}~\bibnamefont
  {Ra’Di}}, \bibinfo {author} {\bibfnamefont {C.}~\bibnamefont {Simovski}},\
  and\ \bibinfo {author} {\bibfnamefont {S.}~\bibnamefont {Tretyakov}},\
  }\bibfield  {title} {\bibinfo {title} {Thin perfect absorbers for
  electromagnetic waves: theory, design, and realizations},\ }\href@noop {}
  {\bibfield  {journal} {\bibinfo  {journal} {Phys. Rev. Appl.}\ }\textbf
  {\bibinfo {volume} {3}},\ \bibinfo {pages} {037001} (\bibinfo {year}
  {2015})}\BibitemShut {NoStop}%
\bibitem [{\citenamefont {Asadchy}\ \emph
  {et~al.}(2015{\natexlab{b}})\citenamefont {Asadchy}, \citenamefont
  {Faniayeu}, \citenamefont {Ra’Di}, \citenamefont {Khakhomov}, \citenamefont
  {Semchenko},\ and\ \citenamefont {Tretyakov}}]{asadchy2015broadband}%
  \BibitemOpen
  \bibfield  {author} {\bibinfo {author} {\bibfnamefont {V.~S.}\ \bibnamefont
  {Asadchy}}, \bibinfo {author} {\bibfnamefont {I.~A.}\ \bibnamefont
  {Faniayeu}}, \bibinfo {author} {\bibfnamefont {Y.}~\bibnamefont {Ra’Di}},
  \bibinfo {author} {\bibfnamefont {S.}~\bibnamefont {Khakhomov}}, \bibinfo
  {author} {\bibfnamefont {I.}~\bibnamefont {Semchenko}},\ and\ \bibinfo
  {author} {\bibfnamefont {S.}~\bibnamefont {Tretyakov}},\ }\bibfield  {title}
  {\bibinfo {title} {Broadband reflectionless metasheets: frequency-selective
  transmission and perfect absorption},\ }\href@noop {} {\bibfield  {journal}
  {\bibinfo  {journal} {Phys. Rev. X}\ }\textbf {\bibinfo {volume} {5}},\
  \bibinfo {pages} {031005} (\bibinfo {year} {2015}{\natexlab{b}})}\BibitemShut
  {NoStop}%
\bibitem [{\citenamefont {Tretyakov}(2003)}]{tretyakov2003analytical}%
  \BibitemOpen
  \bibfield  {author} {\bibinfo {author} {\bibfnamefont {S.}~\bibnamefont
  {Tretyakov}},\ }\href@noop {} {\emph {\bibinfo {title} {Analytical modeling
  in applied electromagnetics}}}\ (\bibinfo  {publisher} {Artech House},\
  \bibinfo {year} {2003})\BibitemShut {NoStop}%
\bibitem [{\citenamefont {Epstein}\ and\ \citenamefont
  {Eleftheriades}(2016{\natexlab{c}})}]{epstein2016huygens}%
  \BibitemOpen
  \bibfield  {author} {\bibinfo {author} {\bibfnamefont {A.}~\bibnamefont
  {Epstein}}\ and\ \bibinfo {author} {\bibfnamefont {G.~V.}\ \bibnamefont
  {Eleftheriades}},\ }\bibfield  {title} {\bibinfo {title} {Huygens’
  metasurfaces via the equivalence principle: design and applications},\
  }\href@noop {} {\bibfield  {journal} {\bibinfo  {journal} {J. Opt. Soc. Am.
  B}\ }\textbf {\bibinfo {volume} {33}},\ \bibinfo {pages} {A31} (\bibinfo
  {year} {2016}{\natexlab{c}})}\BibitemShut {NoStop}%
\bibitem [{\citenamefont {Sell}\ \emph {et~al.}(2017)\citenamefont {Sell},
  \citenamefont {Yang}, \citenamefont {Doshay}, \citenamefont {Yang},\ and\
  \citenamefont {Fan}}]{sell2017large}%
  \BibitemOpen
  \bibfield  {author} {\bibinfo {author} {\bibfnamefont {D.}~\bibnamefont
  {Sell}}, \bibinfo {author} {\bibfnamefont {J.}~\bibnamefont {Yang}}, \bibinfo
  {author} {\bibfnamefont {S.}~\bibnamefont {Doshay}}, \bibinfo {author}
  {\bibfnamefont {R.}~\bibnamefont {Yang}},\ and\ \bibinfo {author}
  {\bibfnamefont {J.~A.}\ \bibnamefont {Fan}},\ }\bibfield  {title} {\bibinfo
  {title} {Large-angle, multifunctional metagratings based on freeform
  multimode geometries},\ }\href@noop {} {\bibfield  {journal} {\bibinfo
  {journal} {Nano Lett.}\ }\textbf {\bibinfo {volume} {17}},\ \bibinfo {pages}
  {3752} (\bibinfo {year} {2017})}\BibitemShut {NoStop}%
\bibitem [{\citenamefont {Khaidarov}\ \emph {et~al.}(2017)\citenamefont
  {Khaidarov}, \citenamefont {Hao}, \citenamefont {Paniagua-Dom{\'\i}nguez},
  \citenamefont {Yu}, \citenamefont {Fu}, \citenamefont {Valuckas},
  \citenamefont {Yap}, \citenamefont {Toh}, \citenamefont {Ng},\ and\
  \citenamefont {Kuznetsov}}]{khaidarov2017asymmetric}%
  \BibitemOpen
  \bibfield  {author} {\bibinfo {author} {\bibfnamefont {E.}~\bibnamefont
  {Khaidarov}}, \bibinfo {author} {\bibfnamefont {H.}~\bibnamefont {Hao}},
  \bibinfo {author} {\bibfnamefont {R.}~\bibnamefont
  {Paniagua-Dom{\'\i}nguez}}, \bibinfo {author} {\bibfnamefont {Y.~F.}\
  \bibnamefont {Yu}}, \bibinfo {author} {\bibfnamefont {Y.~H.}\ \bibnamefont
  {Fu}}, \bibinfo {author} {\bibfnamefont {V.}~\bibnamefont {Valuckas}},
  \bibinfo {author} {\bibfnamefont {S.~L.~K.}\ \bibnamefont {Yap}}, \bibinfo
  {author} {\bibfnamefont {Y.~T.}\ \bibnamefont {Toh}}, \bibinfo {author}
  {\bibfnamefont {J.~S.~K.}\ \bibnamefont {Ng}},\ and\ \bibinfo {author}
  {\bibfnamefont {A.~I.}\ \bibnamefont {Kuznetsov}},\ }\bibfield  {title}
  {\bibinfo {title} {Asymmetric nanoantennas for ultrahigh angle broadband
  visible light bending},\ }\href@noop {} {\bibfield  {journal} {\bibinfo
  {journal} {Nano Lett.}\ }\textbf {\bibinfo {volume} {17}},\ \bibinfo {pages}
  {6267} (\bibinfo {year} {2017})}\BibitemShut {NoStop}%
\bibitem [{\citenamefont {Yang}\ \emph {et~al.}(2018)\citenamefont {Yang},
  \citenamefont {Sell},\ and\ \citenamefont {Fan}}]{yang2018freeform}%
  \BibitemOpen
  \bibfield  {author} {\bibinfo {author} {\bibfnamefont {J.}~\bibnamefont
  {Yang}}, \bibinfo {author} {\bibfnamefont {D.}~\bibnamefont {Sell}},\ and\
  \bibinfo {author} {\bibfnamefont {J.~A.}\ \bibnamefont {Fan}},\ }\bibfield
  {title} {\bibinfo {title} {Freeform metagratings based on complex light
  scattering dynamics for extreme, high efficiency beam steering},\ }\href@noop
  {} {\bibfield  {journal} {\bibinfo  {journal} {Ann. Phys.}\ }\textbf
  {\bibinfo {volume} {530}},\ \bibinfo {pages} {1700302} (\bibinfo {year}
  {2018})}\BibitemShut {NoStop}%
\bibitem [{\citenamefont {Ra’di}\ \emph {et~al.}(2017)\citenamefont
  {Ra’di}, \citenamefont {Sounas},\ and\ \citenamefont
  {Al{\`u}}}]{ra2017metagratings}%
  \BibitemOpen
  \bibfield  {author} {\bibinfo {author} {\bibfnamefont {Y.}~\bibnamefont
  {Ra’di}}, \bibinfo {author} {\bibfnamefont {D.~L.}\ \bibnamefont
  {Sounas}},\ and\ \bibinfo {author} {\bibfnamefont {A.}~\bibnamefont
  {Al{\`u}}},\ }\bibfield  {title} {\bibinfo {title} {Metagratings: beyond the
  limits of graded metasurfaces for wave front control},\ }\href@noop {}
  {\bibfield  {journal} {\bibinfo  {journal} {Phys. Rev. Lett.}\ }\textbf
  {\bibinfo {volume} {119}},\ \bibinfo {pages} {067404} (\bibinfo {year}
  {2017})}\BibitemShut {NoStop}%
\bibitem [{\citenamefont {Memarian}\ \emph {et~al.}(2017)\citenamefont
  {Memarian}, \citenamefont {Li}, \citenamefont {Morimoto},\ and\ \citenamefont
  {Itoh}}]{memarian2017wide}%
  \BibitemOpen
  \bibfield  {author} {\bibinfo {author} {\bibfnamefont {M.}~\bibnamefont
  {Memarian}}, \bibinfo {author} {\bibfnamefont {X.}~\bibnamefont {Li}},
  \bibinfo {author} {\bibfnamefont {Y.}~\bibnamefont {Morimoto}},\ and\
  \bibinfo {author} {\bibfnamefont {T.}~\bibnamefont {Itoh}},\ }\bibfield
  {title} {\bibinfo {title} {Wide-band/angle blazed surfaces using multiple
  coupled blazing resonances},\ }\href@noop {} {\bibfield  {journal} {\bibinfo
  {journal} {Sci. Rep.}\ }\textbf {\bibinfo {volume} {7}} (\bibinfo {year}
  {2017})}\BibitemShut {NoStop}%
\bibitem [{\citenamefont {Chalabi}\ \emph {et~al.}(2017)\citenamefont
  {Chalabi}, \citenamefont {Ra'Di}, \citenamefont {Sounas},\ and\ \citenamefont
  {Al{\`u}}}]{chalabi2017efficient}%
  \BibitemOpen
  \bibfield  {author} {\bibinfo {author} {\bibfnamefont {H.}~\bibnamefont
  {Chalabi}}, \bibinfo {author} {\bibfnamefont {Y.}~\bibnamefont {Ra'Di}},
  \bibinfo {author} {\bibfnamefont {D.}~\bibnamefont {Sounas}},\ and\ \bibinfo
  {author} {\bibfnamefont {A.}~\bibnamefont {Al{\`u}}},\ }\bibfield  {title}
  {\bibinfo {title} {Efficient anomalous reflection through near-field
  interactions in metasurfaces},\ }\href@noop {} {\bibfield  {journal}
  {\bibinfo  {journal} {Phys. Rev. B}\ }\textbf {\bibinfo {volume} {96}},\
  \bibinfo {pages} {075432} (\bibinfo {year} {2017})}\BibitemShut {NoStop}%
\bibitem [{\citenamefont {Epstein}\ and\ \citenamefont
  {Rabinovich}(2017)}]{epstein2017unveiling}%
  \BibitemOpen
  \bibfield  {author} {\bibinfo {author} {\bibfnamefont {A.}~\bibnamefont
  {Epstein}}\ and\ \bibinfo {author} {\bibfnamefont {O.}~\bibnamefont
  {Rabinovich}},\ }\bibfield  {title} {\bibinfo {title} {Unveiling the
  properties of metagratings via a detailed analytical model for synthesis and
  analysis},\ }\href@noop {} {\bibfield  {journal} {\bibinfo  {journal} {Phys.
  Rev. Appl.}\ }\textbf {\bibinfo {volume} {8}},\ \bibinfo {pages} {054037}
  (\bibinfo {year} {2017})}\BibitemShut {NoStop}%
\bibitem [{\citenamefont {Wong}\ and\ \citenamefont
  {Eleftheriades}(2018)}]{wong2018perfect}%
  \BibitemOpen
  \bibfield  {author} {\bibinfo {author} {\bibfnamefont {A.~M.}\ \bibnamefont
  {Wong}}\ and\ \bibinfo {author} {\bibfnamefont {G.~V.}\ \bibnamefont
  {Eleftheriades}},\ }\bibfield  {title} {\bibinfo {title} {Perfect anomalous
  reflection with a bipartite huygens’ metasurface},\ }\href@noop {}
  {\bibfield  {journal} {\bibinfo  {journal} {Phys. Rev. X}\ }\textbf {\bibinfo
  {volume} {8}},\ \bibinfo {pages} {011036} (\bibinfo {year}
  {2018})}\BibitemShut {NoStop}%
\bibitem [{\citenamefont {Collin}(1990)}]{collin1990field}%
  \BibitemOpen
  \bibfield  {author} {\bibinfo {author} {\bibfnamefont {R.~E.}\ \bibnamefont
  {Collin}},\ }\href@noop {} {\emph {\bibinfo {title} {Field theory of guided
  waves}}},\ Vol.~\bibinfo {volume} {5}\ (\bibinfo  {publisher} {John Wiley \&
  Sons},\ \bibinfo {year} {1990})\BibitemShut {NoStop}%
\bibitem [{\citenamefont {Rabinovich}\ and\ \citenamefont
  {Epstein}(2018)}]{Rabinovich2018AnalyticalReflection}%
  \BibitemOpen
  \bibfield  {author} {\bibinfo {author} {\bibfnamefont {O.}~\bibnamefont
  {Rabinovich}}\ and\ \bibinfo {author} {\bibfnamefont {A.}~\bibnamefont
  {Epstein}},\ }\bibfield  {title} {\bibinfo {title} {{Analytical design of
  printed circuit board (PCB) metagratings for perfect anomalous reflection}},\
  }\href {https://doi.org/10.1109/TAP.2018.2836379} {\bibfield  {journal}
  {\bibinfo  {journal} {IEEE Trans. Antennas Propag.}\ }\textbf {\bibinfo
  {volume} {66}},\ \bibinfo {pages} {4086} (\bibinfo {year}
  {2018})}\BibitemShut {NoStop}%
\bibitem [{\citenamefont {Wong}\ \emph {et~al.}(2018)\citenamefont {Wong},
  \citenamefont {Christian},\ and\ \citenamefont
  {Eleftheriades}}]{wong2018binary}%
  \BibitemOpen
  \bibfield  {author} {\bibinfo {author} {\bibfnamefont {A.~M.}\ \bibnamefont
  {Wong}}, \bibinfo {author} {\bibfnamefont {P.}~\bibnamefont {Christian}},\
  and\ \bibinfo {author} {\bibfnamefont {G.~V.}\ \bibnamefont
  {Eleftheriades}},\ }\bibfield  {title} {\bibinfo {title} {Binary huygens’
  metasurfaces: Experimental demonstration of simple and efficient near-grazing
  retroreflectors for te and tm polarizations},\ }\href@noop {} {\bibfield
  {journal} {\bibinfo  {journal} {IEEE Trans. Antennas Propag.}\ }\textbf
  {\bibinfo {volume} {66}},\ \bibinfo {pages} {2892} (\bibinfo {year}
  {2018})}\BibitemShut {NoStop}%
\bibitem [{\citenamefont {Popov}\ \emph {et~al.}(2018)\citenamefont {Popov},
  \citenamefont {Boust},\ and\ \citenamefont {Burokur}}]{popov2018controlling}%
  \BibitemOpen
  \bibfield  {author} {\bibinfo {author} {\bibfnamefont {V.}~\bibnamefont
  {Popov}}, \bibinfo {author} {\bibfnamefont {F.}~\bibnamefont {Boust}},\ and\
  \bibinfo {author} {\bibfnamefont {S.~N.}\ \bibnamefont {Burokur}},\
  }\bibfield  {title} {\bibinfo {title} {Controlling diffraction patterns with
  metagratings},\ }\href@noop {} {\bibfield  {journal} {\bibinfo  {journal}
  {Phys. Rev. Appl.}\ }\textbf {\bibinfo {volume} {10}},\ \bibinfo {pages}
  {011002} (\bibinfo {year} {2018})}\BibitemShut {NoStop}%
\bibitem [{\citenamefont {Neder}\ \emph {et~al.}(2019)\citenamefont {Neder},
  \citenamefont {Ra’di}, \citenamefont {Al{\`u}},\ and\ \citenamefont
  {Polman}}]{neder2019combined}%
  \BibitemOpen
  \bibfield  {author} {\bibinfo {author} {\bibfnamefont {V.}~\bibnamefont
  {Neder}}, \bibinfo {author} {\bibfnamefont {Y.}~\bibnamefont {Ra’di}},
  \bibinfo {author} {\bibfnamefont {A.}~\bibnamefont {Al{\`u}}},\ and\ \bibinfo
  {author} {\bibfnamefont {A.}~\bibnamefont {Polman}},\ }\bibfield  {title}
  {\bibinfo {title} {Combined metagratings for efficient broad-angle scattering
  metasurface},\ }\href@noop {} {\bibfield  {journal} {\bibinfo  {journal} {ACS
  photonics}\ }\textbf {\bibinfo {volume} {6}},\ \bibinfo {pages} {1010}
  (\bibinfo {year} {2019})}\BibitemShut {NoStop}%
\bibitem [{\citenamefont {Popov}\ \emph
  {et~al.}(2019{\natexlab{a}})\citenamefont {Popov}, \citenamefont {Boust},\
  and\ \citenamefont {Burokur}}]{popov2019constructing}%
  \BibitemOpen
  \bibfield  {author} {\bibinfo {author} {\bibfnamefont {V.}~\bibnamefont
  {Popov}}, \bibinfo {author} {\bibfnamefont {F.}~\bibnamefont {Boust}},\ and\
  \bibinfo {author} {\bibfnamefont {S.~N.}\ \bibnamefont {Burokur}},\
  }\bibfield  {title} {\bibinfo {title} {Constructing the near field and far
  field with reactive metagratings: Study on the degrees of freedom},\
  }\href@noop {} {\bibfield  {journal} {\bibinfo  {journal} {Phys. Rev. Appl.}\
  }\textbf {\bibinfo {volume} {11}},\ \bibinfo {pages} {024074} (\bibinfo
  {year} {2019}{\natexlab{a}})}\BibitemShut {NoStop}%
\bibitem [{\citenamefont {Popov}\ \emph
  {et~al.}(2019{\natexlab{b}})\citenamefont {Popov}, \citenamefont {Yakovleva},
  \citenamefont {Boust}, \citenamefont {Pelouard}, \citenamefont {Pardo},\ and\
  \citenamefont {Burokur}}]{popov2019designing}%
  \BibitemOpen
  \bibfield  {author} {\bibinfo {author} {\bibfnamefont {V.}~\bibnamefont
  {Popov}}, \bibinfo {author} {\bibfnamefont {M.}~\bibnamefont {Yakovleva}},
  \bibinfo {author} {\bibfnamefont {F.}~\bibnamefont {Boust}}, \bibinfo
  {author} {\bibfnamefont {J.-L.}\ \bibnamefont {Pelouard}}, \bibinfo {author}
  {\bibfnamefont {F.}~\bibnamefont {Pardo}},\ and\ \bibinfo {author}
  {\bibfnamefont {S.~N.}\ \bibnamefont {Burokur}},\ }\bibfield  {title}
  {\bibinfo {title} {Designing metagratings via local periodic approximation:
  From microwaves to infrared},\ }\href@noop {} {\bibfield  {journal} {\bibinfo
   {journal} {Phys. Rev. Appl.}\ }\textbf {\bibinfo {volume} {11}},\ \bibinfo
  {pages} {044054} (\bibinfo {year} {2019}{\natexlab{b}})}\BibitemShut
  {NoStop}%
\bibitem [{\citenamefont {Sell}\ \emph {et~al.}(2018)\citenamefont {Sell},
  \citenamefont {Yang}, \citenamefont {Wang}, \citenamefont {Phan},
  \citenamefont {Doshay},\ and\ \citenamefont {Fan}}]{sell2018ultra}%
  \BibitemOpen
  \bibfield  {author} {\bibinfo {author} {\bibfnamefont {D.}~\bibnamefont
  {Sell}}, \bibinfo {author} {\bibfnamefont {J.}~\bibnamefont {Yang}}, \bibinfo
  {author} {\bibfnamefont {E.~W.}\ \bibnamefont {Wang}}, \bibinfo {author}
  {\bibfnamefont {T.}~\bibnamefont {Phan}}, \bibinfo {author} {\bibfnamefont
  {S.}~\bibnamefont {Doshay}},\ and\ \bibinfo {author} {\bibfnamefont {J.~A.}\
  \bibnamefont {Fan}},\ }\bibfield  {title} {\bibinfo {title}
  {Ultra-high-efficiency anomalous refraction with dielectric metasurfaces},\
  }\href@noop {} {\bibfield  {journal} {\bibinfo  {journal} {ACS Photonics}\
  }\textbf {\bibinfo {volume} {5}},\ \bibinfo {pages} {2402} (\bibinfo {year}
  {2018})}\BibitemShut {NoStop}%
\bibitem [{\citenamefont {Fan}\ \emph {et~al.}(2018)\citenamefont {Fan},
  \citenamefont {Shcherbakov}, \citenamefont {Allen}, \citenamefont {Allen},
  \citenamefont {Wenner},\ and\ \citenamefont {Shvets}}]{fan2018perfect}%
  \BibitemOpen
  \bibfield  {author} {\bibinfo {author} {\bibfnamefont {Z.}~\bibnamefont
  {Fan}}, \bibinfo {author} {\bibfnamefont {M.~R.}\ \bibnamefont
  {Shcherbakov}}, \bibinfo {author} {\bibfnamefont {M.}~\bibnamefont {Allen}},
  \bibinfo {author} {\bibfnamefont {J.}~\bibnamefont {Allen}}, \bibinfo
  {author} {\bibfnamefont {B.}~\bibnamefont {Wenner}},\ and\ \bibinfo {author}
  {\bibfnamefont {G.}~\bibnamefont {Shvets}},\ }\bibfield  {title} {\bibinfo
  {title} {Perfect diffraction with multiresonant bianisotropic metagratings},\
  }\href@noop {} {\bibfield  {journal} {\bibinfo  {journal} {ACS Photonics}\
  }\textbf {\bibinfo {volume} {5}},\ \bibinfo {pages} {4303} (\bibinfo {year}
  {2018})}\BibitemShut {NoStop}%
\bibitem [{\citenamefont {Casolaro}\ \emph {et~al.}(2019)\citenamefont
  {Casolaro}, \citenamefont {Toscano}, \citenamefont {Al{\`u}},\ and\
  \citenamefont {Bilotti}}]{casolaro2019dynamic}%
  \BibitemOpen
  \bibfield  {author} {\bibinfo {author} {\bibfnamefont {A.}~\bibnamefont
  {Casolaro}}, \bibinfo {author} {\bibfnamefont {A.}~\bibnamefont {Toscano}},
  \bibinfo {author} {\bibfnamefont {A.}~\bibnamefont {Al{\`u}}},\ and\ \bibinfo
  {author} {\bibfnamefont {F.}~\bibnamefont {Bilotti}},\ }\bibfield  {title}
  {\bibinfo {title} {Dynamic beam steering with reconfigurable metagratings},\
  }\href@noop {} {\bibfield  {journal} {\bibinfo  {journal} {IEEE Trans.
  Antennas Propag.}\ } (\bibinfo {year} {2019})}\BibitemShut {NoStop}%
\bibitem [{\citenamefont {{Rabinovich}}\ and\ \citenamefont
  {{Epstein}}(2020)}]{RabinovichArbitrary2020}%
  \BibitemOpen
  \bibfield  {author} {\bibinfo {author} {\bibfnamefont {O.}~\bibnamefont
  {{Rabinovich}}}\ and\ \bibinfo {author} {\bibfnamefont {A.}~\bibnamefont
  {{Epstein}}},\ }\bibfield  {title} {\bibinfo {title} {Arbitrary diffraction
  engineering with multilayered multielement metagratings},\ }\href@noop {}
  {\bibfield  {journal} {\bibinfo  {journal} {IEEE Trans. Antennas Propag.}\
  }\textbf {\bibinfo {volume} {68}},\ \bibinfo {pages} {1553} (\bibinfo {year}
  {2020})}\BibitemShut {NoStop}%
\bibitem [{\citenamefont {Dong}\ \emph {et~al.}(2020)\citenamefont {Dong},
  \citenamefont {Cheng}, \citenamefont {Fan}, \citenamefont {Wang},\ and\
  \citenamefont {Chang}}]{dong2020efficient}%
  \BibitemOpen
  \bibfield  {author} {\bibinfo {author} {\bibfnamefont {X.}~\bibnamefont
  {Dong}}, \bibinfo {author} {\bibfnamefont {J.}~\bibnamefont {Cheng}},
  \bibinfo {author} {\bibfnamefont {F.}~\bibnamefont {Fan}}, \bibinfo {author}
  {\bibfnamefont {X.}~\bibnamefont {Wang}},\ and\ \bibinfo {author}
  {\bibfnamefont {S.}~\bibnamefont {Chang}},\ }\bibfield  {title} {\bibinfo
  {title} {Efficient wide-band large-angle refraction and splitting of a
  terahertz beam by low-index 3d-printed bilayer metagratings},\ }\href@noop {}
  {\bibfield  {journal} {\bibinfo  {journal} {Phys. Rev. Appl.}\ }\textbf
  {\bibinfo {volume} {14}},\ \bibinfo {pages} {014064} (\bibinfo {year}
  {2020})}\BibitemShut {NoStop}%
\bibitem [{\citenamefont {Paniagua-Dominguez}\ \emph
  {et~al.}(2018)\citenamefont {Paniagua-Dominguez}, \citenamefont {Yu},
  \citenamefont {Khaidarov}, \citenamefont {Choi}, \citenamefont {Leong},
  \citenamefont {Bakker}, \citenamefont {Liang}, \citenamefont {Fu},
  \citenamefont {Valuckas}, \citenamefont {Krivitsky} \emph
  {et~al.}}]{paniagua2018metalens}%
  \BibitemOpen
  \bibfield  {author} {\bibinfo {author} {\bibfnamefont {R.}~\bibnamefont
  {Paniagua-Dominguez}}, \bibinfo {author} {\bibfnamefont {Y.~F.}\ \bibnamefont
  {Yu}}, \bibinfo {author} {\bibfnamefont {E.}~\bibnamefont {Khaidarov}},
  \bibinfo {author} {\bibfnamefont {S.}~\bibnamefont {Choi}}, \bibinfo {author}
  {\bibfnamefont {V.}~\bibnamefont {Leong}}, \bibinfo {author} {\bibfnamefont
  {R.~M.}\ \bibnamefont {Bakker}}, \bibinfo {author} {\bibfnamefont
  {X.}~\bibnamefont {Liang}}, \bibinfo {author} {\bibfnamefont {Y.~H.}\
  \bibnamefont {Fu}}, \bibinfo {author} {\bibfnamefont {V.}~\bibnamefont
  {Valuckas}}, \bibinfo {author} {\bibfnamefont {L.~A.}\ \bibnamefont
  {Krivitsky}}, \emph {et~al.},\ }\bibfield  {title} {\bibinfo {title} {A
  metalens with a near-unity numerical aperture},\ }\href@noop {} {\bibfield
  {journal} {\bibinfo  {journal} {Nano Lett.}\ }\textbf {\bibinfo {volume}
  {18}},\ \bibinfo {pages} {2124} (\bibinfo {year} {2018})}\BibitemShut
  {NoStop}%
\bibitem [{\citenamefont {Kang}\ \emph {et~al.}(2020)\citenamefont {Kang},
  \citenamefont {Ra'di}, \citenamefont {Farfan},\ and\ \citenamefont
  {Al{\`u}}}]{kang2020efficient}%
  \BibitemOpen
  \bibfield  {author} {\bibinfo {author} {\bibfnamefont {M.}~\bibnamefont
  {Kang}}, \bibinfo {author} {\bibfnamefont {Y.}~\bibnamefont {Ra'di}},
  \bibinfo {author} {\bibfnamefont {D.}~\bibnamefont {Farfan}},\ and\ \bibinfo
  {author} {\bibfnamefont {A.}~\bibnamefont {Al{\`u}}},\ }\bibfield  {title}
  {\bibinfo {title} {Efficient focusing with large numerical aperture using a
  hybrid metalens},\ }\href@noop {} {\bibfield  {journal} {\bibinfo  {journal}
  {Phys. Rev. Appl.}\ }\textbf {\bibinfo {volume} {13}},\ \bibinfo {pages}
  {044016} (\bibinfo {year} {2020})}\BibitemShut {NoStop}%
\bibitem [{\citenamefont {Elsherbini}\ and\ \citenamefont
  {Sarabandi}(2012)}]{elsherbini2012compact}%
  \BibitemOpen
  \bibfield  {author} {\bibinfo {author} {\bibfnamefont {A.}~\bibnamefont
  {Elsherbini}}\ and\ \bibinfo {author} {\bibfnamefont {K.}~\bibnamefont
  {Sarabandi}},\ }\bibfield  {title} {\bibinfo {title} {Compact directive
  ultra-wideband rectangular waveguide based antenna for radar and
  communication applications},\ }\href@noop {} {\bibfield  {journal} {\bibinfo
  {journal} {IEEE Trans. Antennas Propag.}\ }\textbf {\bibinfo {volume} {60}},\
  \bibinfo {pages} {2203} (\bibinfo {year} {2012})}\BibitemShut {NoStop}%
\bibitem [{\citenamefont {Gonzalez-Ovejero}\ \emph {et~al.}(2018)\citenamefont
  {Gonzalez-Ovejero}, \citenamefont {Chahat}, \citenamefont {Sauleau},
  \citenamefont {Chattopadhyay}, \citenamefont {Maci},\ and\ \citenamefont
  {Ettorre}}]{gonzalez2018additive}%
  \BibitemOpen
  \bibfield  {author} {\bibinfo {author} {\bibfnamefont {D.}~\bibnamefont
  {Gonzalez-Ovejero}}, \bibinfo {author} {\bibfnamefont {N.}~\bibnamefont
  {Chahat}}, \bibinfo {author} {\bibfnamefont {R.}~\bibnamefont {Sauleau}},
  \bibinfo {author} {\bibfnamefont {G.}~\bibnamefont {Chattopadhyay}}, \bibinfo
  {author} {\bibfnamefont {S.}~\bibnamefont {Maci}},\ and\ \bibinfo {author}
  {\bibfnamefont {M.}~\bibnamefont {Ettorre}},\ }\bibfield  {title} {\bibinfo
  {title} {Additive manufactured metal-only modulated metasurface antennas},\
  }\href@noop {} {\bibfield  {journal} {\bibinfo  {journal} {IEEE Trans.
  Antennas Propag.}\ }\textbf {\bibinfo {volume} {66}},\ \bibinfo {pages}
  {6106} (\bibinfo {year} {2018})}\BibitemShut {NoStop}%
\bibitem [{\citenamefont {Campo}\ \emph {et~al.}(2020)\citenamefont {Campo},
  \citenamefont {Carluccio}, \citenamefont {Blanco}, \citenamefont {Litschke},
  \citenamefont {Bruni},\ and\ \citenamefont {Llombart}}]{campo2020wideband}%
  \BibitemOpen
  \bibfield  {author} {\bibinfo {author} {\bibfnamefont {M.~A.}\ \bibnamefont
  {Campo}}, \bibinfo {author} {\bibfnamefont {G.}~\bibnamefont {Carluccio}},
  \bibinfo {author} {\bibfnamefont {D.}~\bibnamefont {Blanco}}, \bibinfo
  {author} {\bibfnamefont {O.}~\bibnamefont {Litschke}}, \bibinfo {author}
  {\bibfnamefont {S.}~\bibnamefont {Bruni}},\ and\ \bibinfo {author}
  {\bibfnamefont {N.}~\bibnamefont {Llombart}},\ }\bibfield  {title} {\bibinfo
  {title} {Wideband circularly polarized antenna with in-lens polarizer for
  high-speed communications},\ }\href@noop {} {\bibfield  {journal} {\bibinfo
  {journal} {IEEE Trans. Antennas Propag.}\ }\textbf {\bibinfo {volume} {69}},\
  \bibinfo {pages} {43} (\bibinfo {year} {2020})}\BibitemShut {NoStop}%
\bibitem [{\citenamefont {Alonso-delPino}\ \emph {et~al.}(2020)\citenamefont
  {Alonso-delPino}, \citenamefont {Bosma}, \citenamefont {Jung-Kubiak},
  \citenamefont {Chattopadhyay},\ and\ \citenamefont
  {Llombart}}]{alonso2020wideband}%
  \BibitemOpen
  \bibfield  {author} {\bibinfo {author} {\bibfnamefont {M.}~\bibnamefont
  {Alonso-delPino}}, \bibinfo {author} {\bibfnamefont {S.}~\bibnamefont
  {Bosma}}, \bibinfo {author} {\bibfnamefont {C.}~\bibnamefont {Jung-Kubiak}},
  \bibinfo {author} {\bibfnamefont {G.}~\bibnamefont {Chattopadhyay}},\ and\
  \bibinfo {author} {\bibfnamefont {N.}~\bibnamefont {Llombart}},\ }\bibfield
  {title} {\bibinfo {title} {Wideband multi-mode leaky-wave feed for scanning
  lens phased array at submillimeter wavelengths},\ }\href@noop {} {\bibfield
  {journal} {\bibinfo  {journal} {IEEE Trans. Terahertz Sci. Technol.}\ }
  (\bibinfo {year} {2020})}\BibitemShut {NoStop}%
\bibitem [{\citenamefont {Boria}\ and\ \citenamefont
  {Gimeno}(2007)}]{boria2007waveguide}%
  \BibitemOpen
  \bibfield  {author} {\bibinfo {author} {\bibfnamefont {V.~E.}\ \bibnamefont
  {Boria}}\ and\ \bibinfo {author} {\bibfnamefont {B.}~\bibnamefont {Gimeno}},\
  }\bibfield  {title} {\bibinfo {title} {Waveguide filters for satellites},\
  }\href@noop {} {\bibfield  {journal} {\bibinfo  {journal} {IEEE Microw.
  Mag.}\ }\textbf {\bibinfo {volume} {8}},\ \bibinfo {pages} {60} (\bibinfo
  {year} {2007})}\BibitemShut {NoStop}%
\bibitem [{Note1()}]{Note1}%
  \BibitemOpen
  \bibinfo {note} {$\protect \mbox {TE}_{mn}$ corresponds to RWG transverse
  electric mode of order $m$ and $n$ along the long and short dimensions of the
  cross section, respectively\cite {pozar2011microwave}.}\BibitemShut {Stop}%
\bibitem [{\citenamefont {Zhao}\ \emph
  {et~al.}(2018{\natexlab{a}})\citenamefont {Zhao}, \citenamefont {Wang},\ and\
  \citenamefont {Deng}}]{zhao2018novel-IEEE}%
  \BibitemOpen
  \bibfield  {author} {\bibinfo {author} {\bibfnamefont {P.}~\bibnamefont
  {Zhao}}, \bibinfo {author} {\bibfnamefont {Q.}~\bibnamefont {Wang}},\ and\
  \bibinfo {author} {\bibfnamefont {J.}~\bibnamefont {Deng}},\ }\bibfield
  {title} {\bibinfo {title} {A novel broadband rectangular waveguide
  $\mbox{TE}_{01}$--$\mbox{TE}_{20}$ mode converter},\ }\href@noop {}
  {\bibfield  {journal} {\bibinfo  {journal} {IEEE Microw. Wirel. Compon.
  Lett.}\ }\textbf {\bibinfo {volume} {28}},\ \bibinfo {pages} {747} (\bibinfo
  {year} {2018}{\natexlab{a}})}\BibitemShut {NoStop}%
\bibitem [{\citenamefont {Belaid}\ \emph {et~al.}(2004)\citenamefont {Belaid},
  \citenamefont {Martinez},\ and\ \citenamefont {Wu}}]{belaid2004mode}%
  \BibitemOpen
  \bibfield  {author} {\bibinfo {author} {\bibfnamefont {M.}~\bibnamefont
  {Belaid}}, \bibinfo {author} {\bibfnamefont {R.}~\bibnamefont {Martinez}},\
  and\ \bibinfo {author} {\bibfnamefont {K.}~\bibnamefont {Wu}},\ }\bibfield
  {title} {\bibinfo {title} {A mode transformer using fin-line array for
  spatial power-combiner applications},\ }\href@noop {} {\bibfield  {journal}
  {\bibinfo  {journal} {IEEE Trans. Microw. Theory Techn.}\ }\textbf {\bibinfo
  {volume} {52}},\ \bibinfo {pages} {1191} (\bibinfo {year}
  {2004})}\BibitemShut {NoStop}%
\bibitem [{\citenamefont {Saad}\ \emph {et~al.}(1977)\citenamefont {Saad},
  \citenamefont {Davies},\ and\ \citenamefont {Davies}}]{saad1977analysis}%
  \BibitemOpen
  \bibfield  {author} {\bibinfo {author} {\bibfnamefont {S.}~\bibnamefont
  {Saad}}, \bibinfo {author} {\bibfnamefont {J.}~\bibnamefont {Davies}},\ and\
  \bibinfo {author} {\bibfnamefont {O.}~\bibnamefont {Davies}},\ }\bibfield
  {title} {\bibinfo {title} {Analysis and design of a circular $\mbox{TE}_{10}$
  mode transducer},\ }\href@noop {} {\bibfield  {journal} {\bibinfo  {journal}
  {IEE J. Microwaves Opt. $\&$ Acoust.}\ }\textbf {\bibinfo {volume} {1}},\
  \bibinfo {pages} {58} (\bibinfo {year} {1977})}\BibitemShut {NoStop}%
\bibitem [{\citenamefont {Yeddulla}\ \emph {et~al.}(2009)\citenamefont
  {Yeddulla}, \citenamefont {Tantawi}, \citenamefont {Guo},\ and\ \citenamefont
  {Dolgashev}}]{yeddulla2009analytical}%
  \BibitemOpen
  \bibfield  {author} {\bibinfo {author} {\bibfnamefont {M.}~\bibnamefont
  {Yeddulla}}, \bibinfo {author} {\bibfnamefont {S.}~\bibnamefont {Tantawi}},
  \bibinfo {author} {\bibfnamefont {J.}~\bibnamefont {Guo}},\ and\ \bibinfo
  {author} {\bibfnamefont {V.}~\bibnamefont {Dolgashev}},\ }\bibfield  {title}
  {\bibinfo {title} {An analytical design and analysis method for a high-power
  circular to rectangular waveguide mode converter and its applications},\
  }\href@noop {} {\bibfield  {journal} {\bibinfo  {journal} {IEEE Trans.
  Microw. Theory Techn.}\ }\textbf {\bibinfo {volume} {57}},\ \bibinfo {pages}
  {1516} (\bibinfo {year} {2009})}\BibitemShut {NoStop}%
\bibitem [{\citenamefont {Liu}\ \emph {et~al.}(2016)\citenamefont {Liu},
  \citenamefont {Wang}, \citenamefont {Pu},\ and\ \citenamefont
  {Luo}}]{liu2016design}%
  \BibitemOpen
  \bibfield  {author} {\bibinfo {author} {\bibfnamefont {G.}~\bibnamefont
  {Liu}}, \bibinfo {author} {\bibfnamefont {Y.}~\bibnamefont {Wang}}, \bibinfo
  {author} {\bibfnamefont {Y.}~\bibnamefont {Pu}},\ and\ \bibinfo {author}
  {\bibfnamefont {Y.}~\bibnamefont {Luo}},\ }\bibfield  {title} {\bibinfo
  {title} {Design and microwave measurement of a novel compact
  $\mbox{TE}_{0n}$/ $\mbox{TE}_{1n'}$ - mode converter},\ }\href@noop {}
  {\bibfield  {journal} {\bibinfo  {journal} {IEEE Trans. Microw. Theory
  Techn.}\ }\textbf {\bibinfo {volume} {64}},\ \bibinfo {pages} {4108}
  (\bibinfo {year} {2016})}\BibitemShut {NoStop}%
\bibitem [{\citenamefont {Wang}\ \emph {et~al.}(2016)\citenamefont {Wang},
  \citenamefont {Wang}, \citenamefont {Liu}, \citenamefont {Shu}, \citenamefont
  {Dong}, \citenamefont {Wang}, \citenamefont {Yan}, \citenamefont {Fu},
  \citenamefont {Yao}, \citenamefont {Luo} \emph {et~al.}}]{wang2016wideband}%
  \BibitemOpen
  \bibfield  {author} {\bibinfo {author} {\bibfnamefont {Y.}~\bibnamefont
  {Wang}}, \bibinfo {author} {\bibfnamefont {L.}~\bibnamefont {Wang}}, \bibinfo
  {author} {\bibfnamefont {G.}~\bibnamefont {Liu}}, \bibinfo {author}
  {\bibfnamefont {G.}~\bibnamefont {Shu}}, \bibinfo {author} {\bibfnamefont
  {K.}~\bibnamefont {Dong}}, \bibinfo {author} {\bibfnamefont {J.}~\bibnamefont
  {Wang}}, \bibinfo {author} {\bibfnamefont {R.}~\bibnamefont {Yan}}, \bibinfo
  {author} {\bibfnamefont {H.}~\bibnamefont {Fu}}, \bibinfo {author}
  {\bibfnamefont {Y.}~\bibnamefont {Yao}}, \bibinfo {author} {\bibfnamefont
  {Y.}~\bibnamefont {Luo}}, \emph {et~al.},\ }\bibfield  {title} {\bibinfo
  {title} {Wideband circular $\mbox{TE}_{21}$ and $\mbox{TE}_{01}$ mode
  converters with same exciting topologies},\ }\href@noop {} {\bibfield
  {journal} {\bibinfo  {journal} {IEEE Trans. Electron Devices}\ }\textbf
  {\bibinfo {volume} {63}},\ \bibinfo {pages} {4088} (\bibinfo {year}
  {2016})}\BibitemShut {NoStop}%
\bibitem [{\citenamefont {Ohana}\ \emph {et~al.}(2016)\citenamefont {Ohana},
  \citenamefont {Desiatov}, \citenamefont {Mazurski},\ and\ \citenamefont
  {Levy}}]{ohana2016dielectric}%
  \BibitemOpen
  \bibfield  {author} {\bibinfo {author} {\bibfnamefont {D.}~\bibnamefont
  {Ohana}}, \bibinfo {author} {\bibfnamefont {B.}~\bibnamefont {Desiatov}},
  \bibinfo {author} {\bibfnamefont {N.}~\bibnamefont {Mazurski}},\ and\
  \bibinfo {author} {\bibfnamefont {U.}~\bibnamefont {Levy}},\ }\bibfield
  {title} {\bibinfo {title} {Dielectric metasurface as a platform for spatial
  mode conversion in nanoscale waveguides},\ }\href@noop {} {\bibfield
  {journal} {\bibinfo  {journal} {Nano Lett.}\ }\textbf {\bibinfo {volume}
  {16}},\ \bibinfo {pages} {7956} (\bibinfo {year} {2016})}\BibitemShut
  {NoStop}%
\bibitem [{\citenamefont {Li}\ \emph {et~al.}(2017)\citenamefont {Li},
  \citenamefont {Kim}, \citenamefont {Wang}, \citenamefont {Han}, \citenamefont
  {Shrestha}, \citenamefont {Overvig}, \citenamefont {Lu}, \citenamefont
  {Stein}, \citenamefont {Agarwal}, \citenamefont {Lon{\v{c}}ar} \emph
  {et~al.}}]{li2017controlling}%
  \BibitemOpen
  \bibfield  {author} {\bibinfo {author} {\bibfnamefont {Z.}~\bibnamefont
  {Li}}, \bibinfo {author} {\bibfnamefont {M.-H.}\ \bibnamefont {Kim}},
  \bibinfo {author} {\bibfnamefont {C.}~\bibnamefont {Wang}}, \bibinfo {author}
  {\bibfnamefont {Z.}~\bibnamefont {Han}}, \bibinfo {author} {\bibfnamefont
  {S.}~\bibnamefont {Shrestha}}, \bibinfo {author} {\bibfnamefont {A.~C.}\
  \bibnamefont {Overvig}}, \bibinfo {author} {\bibfnamefont {M.}~\bibnamefont
  {Lu}}, \bibinfo {author} {\bibfnamefont {A.}~\bibnamefont {Stein}}, \bibinfo
  {author} {\bibfnamefont {A.~M.}\ \bibnamefont {Agarwal}}, \bibinfo {author}
  {\bibfnamefont {M.}~\bibnamefont {Lon{\v{c}}ar}}, \emph {et~al.},\ }\bibfield
   {title} {\bibinfo {title} {Controlling propagation and coupling of waveguide
  modes using phase-gradient metasurfaces},\ }\href@noop {} {\bibfield
  {journal} {\bibinfo  {journal} {Nat. Nanotechnol.}\ }\textbf {\bibinfo
  {volume} {12}},\ \bibinfo {pages} {675} (\bibinfo {year} {2017})}\BibitemShut
  {NoStop}%
\bibitem [{\citenamefont {Wang}\ \emph {et~al.}(2019)\citenamefont {Wang},
  \citenamefont {Zhang}, \citenamefont {He}, \citenamefont {Zhu}, \citenamefont
  {Sun},\ and\ \citenamefont {Su}}]{wang2019compact}%
  \BibitemOpen
  \bibfield  {author} {\bibinfo {author} {\bibfnamefont {H.}~\bibnamefont
  {Wang}}, \bibinfo {author} {\bibfnamefont {Y.}~\bibnamefont {Zhang}},
  \bibinfo {author} {\bibfnamefont {Y.}~\bibnamefont {He}}, \bibinfo {author}
  {\bibfnamefont {Q.}~\bibnamefont {Zhu}}, \bibinfo {author} {\bibfnamefont
  {L.}~\bibnamefont {Sun}},\ and\ \bibinfo {author} {\bibfnamefont
  {Y.}~\bibnamefont {Su}},\ }\bibfield  {title} {\bibinfo {title} {Compact
  silicon waveguide mode converter employing dielectric metasurface
  structure},\ }\href@noop {} {\bibfield  {journal} {\bibinfo  {journal} {Adv.
  Opt. Mater.}\ }\textbf {\bibinfo {volume} {7}},\ \bibinfo {pages} {1801191}
  (\bibinfo {year} {2019})}\BibitemShut {NoStop}%
\bibitem [{\citenamefont {Kirilenko}\ \emph {et~al.}(2006)\citenamefont
  {Kirilenko}, \citenamefont {Rud},\ and\ \citenamefont
  {Tkachenko}}]{kirilenko2006nonsymmetrical}%
  \BibitemOpen
  \bibfield  {author} {\bibinfo {author} {\bibfnamefont {A.~A.}\ \bibnamefont
  {Kirilenko}}, \bibinfo {author} {\bibfnamefont {L.~A.}\ \bibnamefont {Rud}},\
  and\ \bibinfo {author} {\bibfnamefont {V.~I.}\ \bibnamefont {Tkachenko}},\
  }\bibfield  {title} {\bibinfo {title} {Nonsymmetrical h-plane corners for
  $\mbox{TE}_{10}$-$\mbox{TE}_{q0}$ mode conversion in rectangular
  waveguides},\ }\href@noop {} {\bibfield  {journal} {\bibinfo  {journal} {IEEE
  Trans. Microw. Theory Techn.}\ }\textbf {\bibinfo {volume} {54}},\ \bibinfo
  {pages} {2471} (\bibinfo {year} {2006})}\BibitemShut {NoStop}%
\bibitem [{\citenamefont {Zhang}\ \emph {et~al.}(2012)\citenamefont {Zhang},
  \citenamefont {Yuan},\ and\ \citenamefont {Liu}}]{zhang2012theoretical}%
  \BibitemOpen
  \bibfield  {author} {\bibinfo {author} {\bibfnamefont {Q.}~\bibnamefont
  {Zhang}}, \bibinfo {author} {\bibfnamefont {C.-W.}\ \bibnamefont {Yuan}},\
  and\ \bibinfo {author} {\bibfnamefont {L.}~\bibnamefont {Liu}},\ }\bibfield
  {title} {\bibinfo {title} {Theoretical design and analysis for
  $\mbox{TE}_{20}$–$\mbox{TE}_{10}$ rectangular waveguide mode converters},\
  }\href@noop {} {\bibfield  {journal} {\bibinfo  {journal} {IEEE Trans.
  Microw. Theory Techn.}\ }\textbf {\bibinfo {volume} {60}},\ \bibinfo {pages}
  {1018} (\bibinfo {year} {2012})}\BibitemShut {NoStop}%
\bibitem [{\citenamefont {Zhao}\ \emph
  {et~al.}(2018{\natexlab{b}})\citenamefont {Zhao}, \citenamefont {Wang},\ and\
  \citenamefont {Deng}}]{zhao2018novel-AIP}%
  \BibitemOpen
  \bibfield  {author} {\bibinfo {author} {\bibfnamefont {P.}~\bibnamefont
  {Zhao}}, \bibinfo {author} {\bibfnamefont {Q.}~\bibnamefont {Wang}},\ and\
  \bibinfo {author} {\bibfnamefont {J.}~\bibnamefont {Deng}},\ }\bibfield
  {title} {\bibinfo {title} {A novel lightweight
  $\mbox{TE}_{01}$-$\mbox{TE}_{20}$ mode converter with broad bandwidth and
  aligned ports},\ }\href@noop {} {\bibfield  {journal} {\bibinfo  {journal}
  {Rev. Sci. Instrum.}\ }\textbf {\bibinfo {volume} {89}},\ \bibinfo {pages}
  {094702} (\bibinfo {year} {2018}{\natexlab{b}})}\BibitemShut {NoStop}%
\bibitem [{\citenamefont {Wu}\ \emph {et~al.}(2013)\citenamefont {Wu},
  \citenamefont {Li}, \citenamefont {Fu}, \citenamefont {Li},\ and\
  \citenamefont {Xu}}]{wu2013te01}%
  \BibitemOpen
  \bibfield  {author} {\bibinfo {author} {\bibfnamefont {Z.}~\bibnamefont
  {Wu}}, \bibinfo {author} {\bibfnamefont {H.}~\bibnamefont {Li}}, \bibinfo
  {author} {\bibfnamefont {H.}~\bibnamefont {Fu}}, \bibinfo {author}
  {\bibfnamefont {T.}~\bibnamefont {Li}},\ and\ \bibinfo {author}
  {\bibfnamefont {J.}~\bibnamefont {Xu}},\ }\bibfield  {title} {\bibinfo
  {title} {A $\mbox{TE}_{01}$ mode generator for testing high power
  transmission devices},\ }\href@noop {} {\bibfield  {journal} {\bibinfo
  {journal} {Rev. Sci. Instrum.}\ }\textbf {\bibinfo {volume} {84}},\ \bibinfo
  {pages} {114702} (\bibinfo {year} {2013})}\BibitemShut {NoStop}%
\bibitem [{\citenamefont {Shu}\ \emph {et~al.}(2020{\natexlab{a}})\citenamefont
  {Shu}, \citenamefont {Cai}, \citenamefont {Li}, \citenamefont {Liu},\ and\
  \citenamefont {He}}]{shu2020wideband}%
  \BibitemOpen
  \bibfield  {author} {\bibinfo {author} {\bibfnamefont {G.}~\bibnamefont
  {Shu}}, \bibinfo {author} {\bibfnamefont {Z.}~\bibnamefont {Cai}}, \bibinfo
  {author} {\bibfnamefont {Y.}~\bibnamefont {Li}}, \bibinfo {author}
  {\bibfnamefont {G.}~\bibnamefont {Liu}},\ and\ \bibinfo {author}
  {\bibfnamefont {W.}~\bibnamefont {He}},\ }\bibfield  {title} {\bibinfo
  {title} {Wideband rectangular $\mbox{TE}_{10}$ to $\mbox{TE}_{n0}$ mode
  converters for terahertz-band high-order overmoded planar slow-wave
  structures},\ }\href@noop {} {\bibfield  {journal} {\bibinfo  {journal} {IEEE
  Trans. Electron Devices}\ }\textbf {\bibinfo {volume} {67}},\ \bibinfo
  {pages} {1259} (\bibinfo {year} {2020}{\natexlab{a}})}\BibitemShut {NoStop}%
\bibitem [{\citenamefont {Xu}\ \emph {et~al.}(2019)\citenamefont {Xu},
  \citenamefont {Peng}, \citenamefont {Sun}, \citenamefont {Luo}, \citenamefont
  {Wang}, \citenamefont {Jiang}, \citenamefont {Liu},\ and\ \citenamefont
  {Wu}}]{xu2019design}%
  \BibitemOpen
  \bibfield  {author} {\bibinfo {author} {\bibfnamefont {Y.}~\bibnamefont
  {Xu}}, \bibinfo {author} {\bibfnamefont {T.}~\bibnamefont {Peng}}, \bibinfo
  {author} {\bibfnamefont {M.}~\bibnamefont {Sun}}, \bibinfo {author}
  {\bibfnamefont {Y.}~\bibnamefont {Luo}}, \bibinfo {author} {\bibfnamefont
  {J.}~\bibnamefont {Wang}}, \bibinfo {author} {\bibfnamefont {W.}~\bibnamefont
  {Jiang}}, \bibinfo {author} {\bibfnamefont {G.}~\bibnamefont {Liu}},\ and\
  \bibinfo {author} {\bibfnamefont {Z.}~\bibnamefont {Wu}},\ }\bibfield
  {title} {\bibinfo {title} {Design and test of broadband rectangular waveguide
  $\mbox{TE}_{10}$ to circular waveguide $\mbox{TE}_{21}$ and $\mbox{TE}_{01}$
  mode converters},\ }\href@noop {} {\bibfield  {journal} {\bibinfo  {journal}
  {IEEE Trans. Electron Devices}\ }\textbf {\bibinfo {volume} {66}},\ \bibinfo
  {pages} {3573} (\bibinfo {year} {2019})}\BibitemShut {NoStop}%
\bibitem [{\citenamefont {Shu}\ \emph {et~al.}(2020{\natexlab{b}})\citenamefont
  {Shu}, \citenamefont {Qian},\ and\ \citenamefont {He}}]{shu2020design}%
  \BibitemOpen
  \bibfield  {author} {\bibinfo {author} {\bibfnamefont {G.}~\bibnamefont
  {Shu}}, \bibinfo {author} {\bibfnamefont {Z.}~\bibnamefont {Qian}},\ and\
  \bibinfo {author} {\bibfnamefont {W.}~\bibnamefont {He}},\ }\bibfield
  {title} {\bibinfo {title} {Design and measurement of an h-band rectangular
  $\mbox{TE}_{10}$ to $\mbox{TE}_{20}$ mode converter},\ }\href@noop {}
  {\bibfield  {journal} {\bibinfo  {journal} {IEEE Access}\ }\textbf {\bibinfo
  {volume} {8}},\ \bibinfo {pages} {37242} (\bibinfo {year}
  {2020}{\natexlab{b}})}\BibitemShut {NoStop}%
\bibitem [{\citenamefont {Harrington}(1961)}]{harrington2001time}%
  \BibitemOpen
  \bibfield  {author} {\bibinfo {author} {\bibfnamefont {R.~F.}\ \bibnamefont
  {Harrington}},\ }\href@noop {} {\bibinfo {title} {Time-harmonic
  electromagnetic fields}} (\bibinfo {year} {1961})\BibitemShut {NoStop}%
\bibitem [{\citenamefont {Ikonen}\ \emph {et~al.}(2007)\citenamefont {Ikonen},
  \citenamefont {Saenz}, \citenamefont {Gonzalo},\ and\ \citenamefont
  {Tretyakov}}]{ikonen2007modeling}%
  \BibitemOpen
  \bibfield  {author} {\bibinfo {author} {\bibfnamefont {P.~M.}\ \bibnamefont
  {Ikonen}}, \bibinfo {author} {\bibfnamefont {E.}~\bibnamefont {Saenz}},
  \bibinfo {author} {\bibfnamefont {R.}~\bibnamefont {Gonzalo}},\ and\ \bibinfo
  {author} {\bibfnamefont {S.~A.}\ \bibnamefont {Tretyakov}},\ }\bibfield
  {title} {\bibinfo {title} {Modeling and analysis of composite antenna
  superstrates consisting on grids of loaded wires},\ }\href@noop {} {\bibfield
   {journal} {\bibinfo  {journal} {IEEE Trans. Antennas Propag.}\ }\textbf
  {\bibinfo {volume} {55}},\ \bibinfo {pages} {2692} (\bibinfo {year}
  {2007})}\BibitemShut {NoStop}%
\bibitem [{\citenamefont {Lewin}(1975)}]{lewin1975theory}%
  \BibitemOpen
  \bibfield  {author} {\bibinfo {author} {\bibfnamefont {L.}~\bibnamefont
  {Lewin}},\ }\bibfield  {title} {\bibinfo {title} {Theory of waveguides:
  Techniques for the solution of waveguide problems},\ }\href@noop {} {\
  (\bibinfo {year} {1975})}\BibitemShut {NoStop}%
\bibitem [{\citenamefont {Leviatan}\ \emph {et~al.}(1983)\citenamefont
  {Leviatan}, \citenamefont {Li}, \citenamefont {Adams},\ and\ \citenamefont
  {Perini}}]{leviatan1983single}%
  \BibitemOpen
  \bibfield  {author} {\bibinfo {author} {\bibfnamefont {Y.}~\bibnamefont
  {Leviatan}}, \bibinfo {author} {\bibfnamefont {P.~G.}\ \bibnamefont {Li}},
  \bibinfo {author} {\bibfnamefont {A.~T.}\ \bibnamefont {Adams}},\ and\
  \bibinfo {author} {\bibfnamefont {J.}~\bibnamefont {Perini}},\ }\bibfield
  {title} {\bibinfo {title} {Single-post inductive obstacle in rectangular
  waveguide},\ }\href@noop {} {\bibfield  {journal} {\bibinfo  {journal} {IEEE
  Trans. Microw. Theory Techn.}\ }\textbf {\bibinfo {volume} {31}},\ \bibinfo
  {pages} {806} (\bibinfo {year} {1983})}\BibitemShut {NoStop}%
\bibitem [{\citenamefont {Rabinovich}\ \emph {et~al.}(2019)\citenamefont
  {Rabinovich}, \citenamefont {Kaplon}, \citenamefont {Reis},\ and\
  \citenamefont {Epstein}}]{rabinovich2019experimental}%
  \BibitemOpen
  \bibfield  {author} {\bibinfo {author} {\bibfnamefont {O.}~\bibnamefont
  {Rabinovich}}, \bibinfo {author} {\bibfnamefont {I.}~\bibnamefont {Kaplon}},
  \bibinfo {author} {\bibfnamefont {J.}~\bibnamefont {Reis}},\ and\ \bibinfo
  {author} {\bibfnamefont {A.}~\bibnamefont {Epstein}},\ }\bibfield  {title}
  {\bibinfo {title} {Experimental demonstration and in-depth investigation of
  analytically designed anomalous reflection metagratings},\ }\href@noop {}
  {\bibfield  {journal} {\bibinfo  {journal} {Phys. Rev. B}\ }\textbf {\bibinfo
  {volume} {99}},\ \bibinfo {pages} {125101} (\bibinfo {year}
  {2019})}\BibitemShut {NoStop}%
\bibitem [{\citenamefont {RS}(1979)}]{retangularwaveguidestandards}%
  \BibitemOpen
  \bibfield  {author} {\bibinfo {author} {\bibfnamefont {E.}~\bibnamefont
  {RS}},\ }\bibfield  {title} {\bibinfo {title} {261-b-rectangular waveguides
  (wr3 to wr2300)},\ }\href@noop {} {\bibfield  {journal} {\bibinfo  {journal}
  {Electronic Industries Association of the United States of America}\ }
  (\bibinfo {year} {1979})}\BibitemShut {NoStop}%
\bibitem [{Note2()}]{Note2}%
  \BibitemOpen
  \bibinfo {note} {In case inductive loading is required, other PCB-compatible
  geometries, such as meander lines, may be used \cite
  {popov2019designing}.}\BibitemShut {Stop}%
\bibitem [{Note3()}]{Note3}%
  \BibitemOpen
  \bibinfo {note} {The correction factor $K_{\protect \mathrm {corr}}$ is
  assessed with the aid of a full-wave simulation. For a chosen reference case
  [configuration $\#10$ of Fig. \ref {fig:Widths of the capacitor}(a) in our
  case, we find the optimal capacitor width in Ansys HFSS and use this value to
  calibrate $K_{\protect \mathrm {corr}}$ such that Eq. \protect \textup {\hbox
  {\mathsurround \z@ \protect \normalfont (\ignorespaces \ref {eq:Width}\unskip
  \@@italiccorr )}} would yield the same result \cite
  {epstein2017unveiling}.}\BibitemShut {Stop}%
\bibitem [{\citenamefont {Balanis}(2012)}]{balanis2012advanced}%
  \BibitemOpen
  \bibfield  {author} {\bibinfo {author} {\bibfnamefont {C.~A.}\ \bibnamefont
  {Balanis}},\ }\href@noop {} {\emph {\bibinfo {title} {Advanced Engineering
  Electromagnetics}}}\ (\bibinfo  {publisher} {{John Wiley \& Sons}},\ \bibinfo
  {address} {Hoboken, NJ},\ \bibinfo {year} {2012})\BibitemShut {NoStop}%
\bibitem [{\citenamefont {Matthaei}\ \emph {et~al.}(1980)\citenamefont
  {Matthaei}, \citenamefont {Young},\ and\ \citenamefont
  {Jones}}]{matthaei1980microwave}%
  \BibitemOpen
  \bibfield  {author} {\bibinfo {author} {\bibfnamefont {G.~L.}\ \bibnamefont
  {Matthaei}}, \bibinfo {author} {\bibfnamefont {L.}~\bibnamefont {Young}},\
  and\ \bibinfo {author} {\bibfnamefont {E.~M.~T.}\ \bibnamefont {Jones}},\
  }\href@noop {} {\emph {\bibinfo {title} {Microwave filters,
  impedance-matching networks, and coupling structures}}}\ (\bibinfo
  {publisher} {Artech house},\ \bibinfo {year} {1980})\BibitemShut {NoStop}%
\bibitem [{\citenamefont {Maaskant}\ and\ \citenamefont
  {Rosen}(2016)}]{maaskant2016teaching}%
  \BibitemOpen
  \bibfield  {author} {\bibinfo {author} {\bibfnamefont {R.}~\bibnamefont
  {Maaskant}}\ and\ \bibinfo {author} {\bibfnamefont {A.}~\bibnamefont
  {Rosen}},\ }\bibfield  {title} {\bibinfo {title} {Teaching and learning
  electromagnetics: An analytical problem-solving approach [education
  corner]},\ }\href@noop {} {\bibfield  {journal} {\bibinfo  {journal} {IEEE
  Antennas Propag. Mag.}\ }\textbf {\bibinfo {volume} {58}},\ \bibinfo {pages}
  {75} (\bibinfo {year} {2016})}\BibitemShut {NoStop}%
\bibitem [{Note4()}]{Note4}%
  \BibitemOpen
  \bibinfo {note} {It is important to include the flange at the output (port 2)
  of the RWG [Fig. \ref {fig:Fabrication prototype}(e)] in full-wave
  simulations ; otherwise, increased backscattering effects will be recorded,
  inconsistent with the measured scenario. Consequently, we have used the
  following dimensions for the flange, extracted from the fabricated prototype,
  in our simulations: the width of the metal rim along $x$ and $y$ are taken as
  $9.1$ mm and $15.45$ mm, respectively, while the thickness along the $z$-axis
  is kept as $1.27$ mm}\BibitemShut {NoStop}%
\bibitem [{\citenamefont {Abramowitz}\ and\ \citenamefont
  {Stegun}(1970)}]{abramowitz1970handbook}%
  \BibitemOpen
  \bibfield  {author} {\bibinfo {author} {\bibfnamefont {M.}~\bibnamefont
  {Abramowitz}}\ and\ \bibinfo {author} {\bibfnamefont {I.~A.}\ \bibnamefont
  {Stegun}},\ }\href@noop {} {\emph {\bibinfo {title} {Handbook of Mathematical
  Functions : with Formulas, Graphs, and Mathematical Tables}}}\ (\bibinfo
  {publisher} {Dover Publications},\ \bibinfo {address} {New York},\ \bibinfo
  {year} {1970})\BibitemShut {NoStop}%
\bibitem [{\citenamefont {Jeffrey}\ and\ \citenamefont
  {Zwillinger}(2007)}]{jeffrey2007table}%
  \BibitemOpen
  \bibfield  {author} {\bibinfo {author} {\bibfnamefont {A.}~\bibnamefont
  {Jeffrey}}\ and\ \bibinfo {author} {\bibfnamefont {D.}~\bibnamefont
  {Zwillinger}},\ }\href@noop {} {\emph {\bibinfo {title} {Table of integrals,
  series, and products}}}\ (\bibinfo  {publisher} {Elsevier},\ \bibinfo {year}
  {2007})\BibitemShut {NoStop}%
\bibitem [{\citenamefont {Pozar}(2011)}]{pozar2011microwave}%
  \BibitemOpen
  \bibfield  {author} {\bibinfo {author} {\bibfnamefont {D.~M.}\ \bibnamefont
  {Pozar}},\ }\href@noop {} {\emph {\bibinfo {title} {Microwave engineering}}}\
  (\bibinfo  {publisher} {John wiley \& sons},\ \bibinfo {year}
  {2011})\BibitemShut {NoStop}%
\end{thebibliography}

\providecommand{\noopsort}[1]{}\providecommand{\singleletter}[1]{#1}%

\end{document}